\documentstyle[multicol,aps,epsf,here]{revtex}
\tighten
\newcommand{\PostScript}[7]{
\begin{figure}[H]
\begin{center}
\leavevmode
\epsfysize=#1cm
\vspace{#2cm}
\epsfbox{#3}
\par
\parbox{#5cm}{
\vspace{#4cm}
\caption[figure]{\renewcommand{\baselinestretch}{1} \small \normalsize #6}
\label{#7}}
\end{center}
\end{figure}
}

\begin{document}

\newcommand{\half}{\frac {1}{2} }
\newcommand{\eg}{{\em e.g.} }
\newcommand{\ie}{{\em i.e.} }
\newcommand{\etc}{{\em etc.}}
\newcommand{\si}{\simeq}
\newcommand{\etal}{{\em et al.\ }}
\newcommand{\cf}{{\em cf. }}

\newcommand{\dd}[2]{{\rmd{#1}\over\rmd{#2}}}
\newcommand{\pdd}[2]{{\partial{#1}\over\partial{#2}}}
\newcommand{\pa}[1]{\partial_{#1}}
\newcommand{\pref}[1]{(\ref{#1})}

\newcommand {\be}[1]{
%%      For draft    %%%%%%%%
%{\marginpar{{\scriptsize\ \\ \ #1}}}
      \begin{eqnarray} \mbox{$\label{#1}$}}

\newcommand{\ee}{\end{eqnarray}}

\def\rdown{\rho_{\downarrow}}
\def\pa{\partial}
\def\th{\theta}
\def\bea{\begin{eqnarray}}
\def\eea{\end{eqnarray}}
\def\be{\begin{equation}}
\def\ee{\end{equation}}
\def\pa{\partial}
\def\eps{\epsilon}
\def\th{\theta}
\def\na{\nabla}
\def\nn{\nonumber}
\def\lan{\langle}
\def\ran{\rangle}
\def\pr{\prime}
\def\rarrow{\rightarrow}
\def\larrow{\leftarrow}

%%%%%%%%%%%%%%%%%%%%%
%\vspace* {-21 mm}
%\begin{flushright}
%  NORDITA-2000/XX CM \\
%  April-2000 \\
%\end{flushright}
%%%%%%%%%%%%%%%%%%%%

\title{Scattering phase shifts in quasi-one-dimension}
\vskip 20 mm
\author{P. Singha Deo and Swarnali Bandopadhyay}
\address{S.N. Bose National Centre for Basic Sciences,
         J.D.Block, Sector III, Salt Lake City, Calcutta 700098, India}
\author{Sourin Das}
\address{Harish Chandra Research Institute, Chhatnag Road, Jhusi,
     Allahabad 211019, India }
\date{\today}

\maketitle

\begin{abstract}

Scattering of an electron in quasi-one dimensional quantum wires have many
unusual features, not found in one, two or three dimensions. In this work
we analyze the scattering phase shifts due to
an impurity in a multi-channel quantum wire
with special emphasis on negative slopes in the scattering phase shift
versus incident energy curves and the Wigner delay time.
Although at first sight, the large number
of scattering matrix elements show phase shifts of different character and
nature, it is possible to see some pattern and understand these features.
The behavior of scattering phase shifts in one-dimension can be seen as a
special case of these features observed in quasi-one-dimensions.
The negative slopes can occur at any arbitrary energy and
Friedel sum rule is completely violated in
quasi-one-dimension at any arbitrary energy and any arbitrary regime.
This is in contrast to one, two or
three dimensions where such negative slopes and violation of
Friedel sum rule happen only at low energy where
the incident electron feels the potential very strongly (i.e.,
there is a very well defined regime, the WKB regime, where
FSR works very well).
There are some novel
behavior of scattering phase shifts at the critical energies where
$S$-matrix changes dimension.\\

\hfill
\end{abstract}
\begin{multicols}{2}

\narrowtext
%%%%%%%%%%%%%%%%%%%%%%
\section{Introduction}
\label{s1}
%%%%%%%%%%%%%%%%%%%%%%

Elastic scattering in one, two and three dimensions is well understood
\cite{new82}. In a scattering process there are some important physical
quantities like scattering amplitude and scattering phase shift. While the
scattering intensity is directly related to the scattering amplitude, the
scattering phase shifts are also very important physical quantities and
the Friedel-sum-rule (FSR) relates them to the density of states (DOS).

%%%
\PostScript{6}{0}{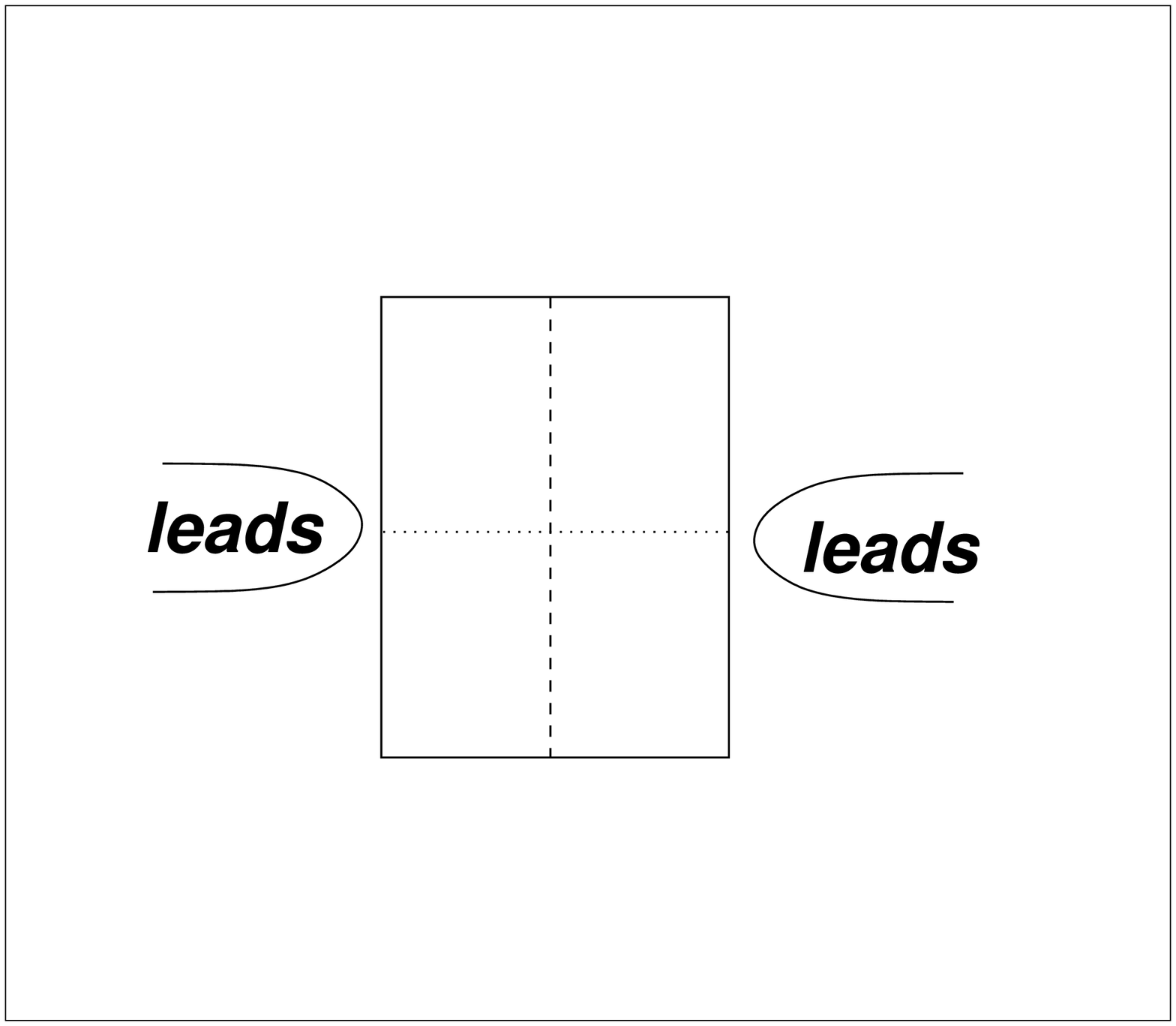}{0.5}{14}{
\hspace{-.5cm}
A rectangular quantum biliard or quantum dot, weakly coupled to leads.
The dotted line is along the x-axis and the dashed line is along
the y-axis.
}{f2}
%%%

At low temperatures, inelastic collisions are greatly suppressed.  As a
result the phase coherence length of an electron can become a few microns.
Mesoscopic systems are defined as systems in which the phase coherence
length exceeds the sample size. In such a system elastic scattering is the
dominant feature and such mesoscopic samples can be understood as phase
coherent elastic scatterers.  With the experimental realizations of
mesoscopic systems and their possibility of being applied in
nano-technology and quantum computing, understanding scattering effects in
quasi-one dimensions (Q1D) has also become important at present. This is
also essential because the Landauer conductance formula relates the
conductance to partial scattering intensities and also one can now probe
scattering phase shift directly in an experiment \cite{yac95,sch97,CER97}.
Mesoscopic samples are normally made up of metals or semiconductors, and
the defects in them are generally point defects. Hence we will restrict
our analysis to delta function potential impurities.

Recently a new kind of scattering phase shift was discussed in Q1D, in
connection with the violation of the parity effect \cite{deo96}. To
explain this phase shift and the violation of parity effect, here we
elaborate some portions of Ref. \cite{deo96} and explain what are symmetry
dictated nodes (SDN) and non-symmetry dictated nodes (NSDN) that can arise
in a Q1D system. For example let us consider a rectangular quantum
biliard or dot connected to leads by quantum mechanical tunneling as
shown in Fig.1.  The system has reflection symmetry across the x-axis as
well as the y-axis. Also the x-y components separate and spanning nodes
(nodes that span across the direction of propagation and shown by dashed
line in Fig.1) as well as non-spanning nodes (shown by dotted line in
Fig.1)  develop in the geometry, dictated by the reflection symmetries.
There are various symmetries that give rise to nodes in the wave-function
and we call them SDN (for example the antisymmetric property of the many
body wave function result in nodes). But if quantum coherence extends to
some distance inside the leads then one can model the phase coherent
quantum dot as shown in Fig.2.

%%%
\PostScript{6}{0}{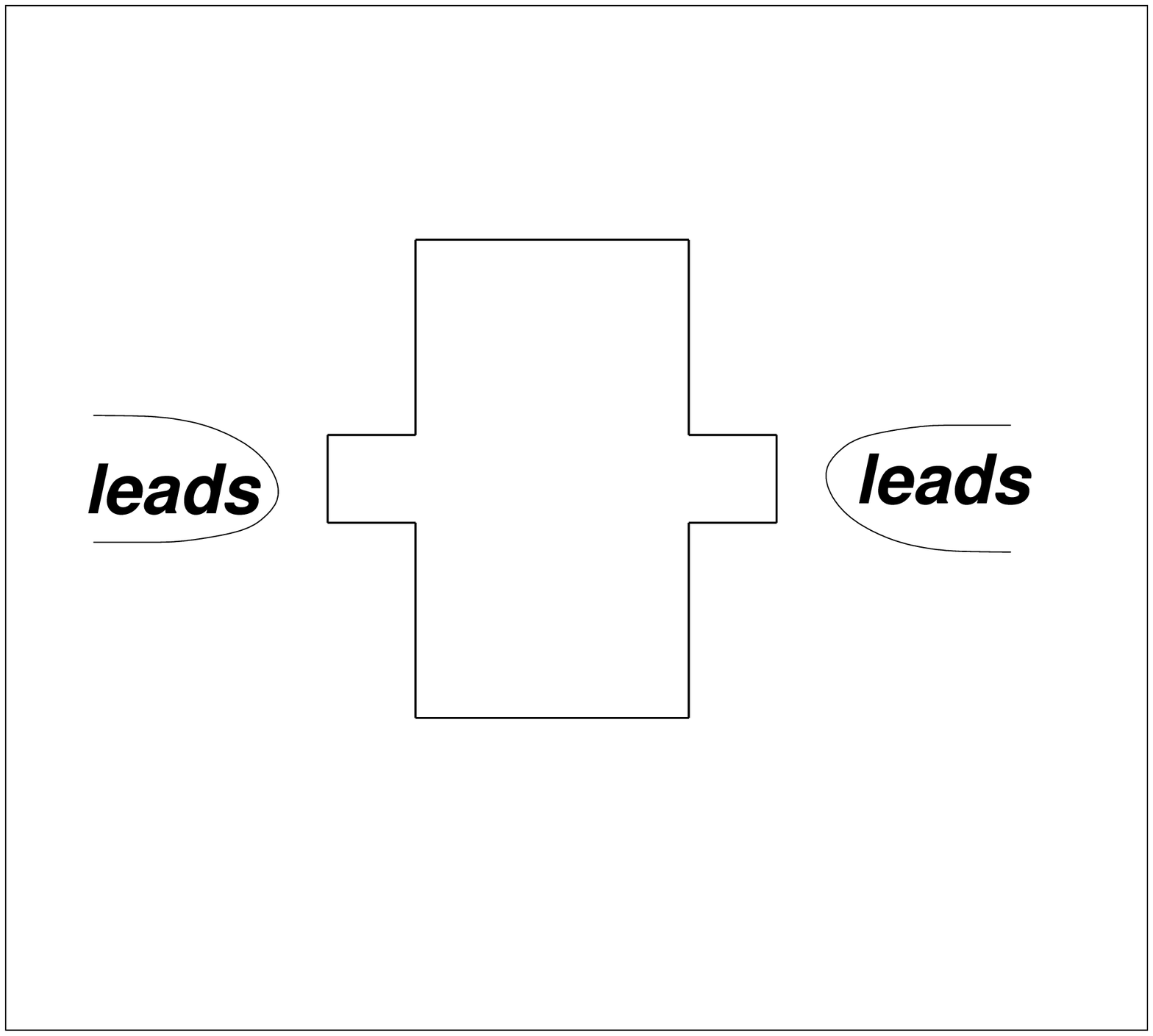}{0.5}{14}{
\hspace{-.5cm}
A more realistic model of the quantum dot of Fig.1.
}{f2}
%%%

%%%
\PostScript{6}{0}{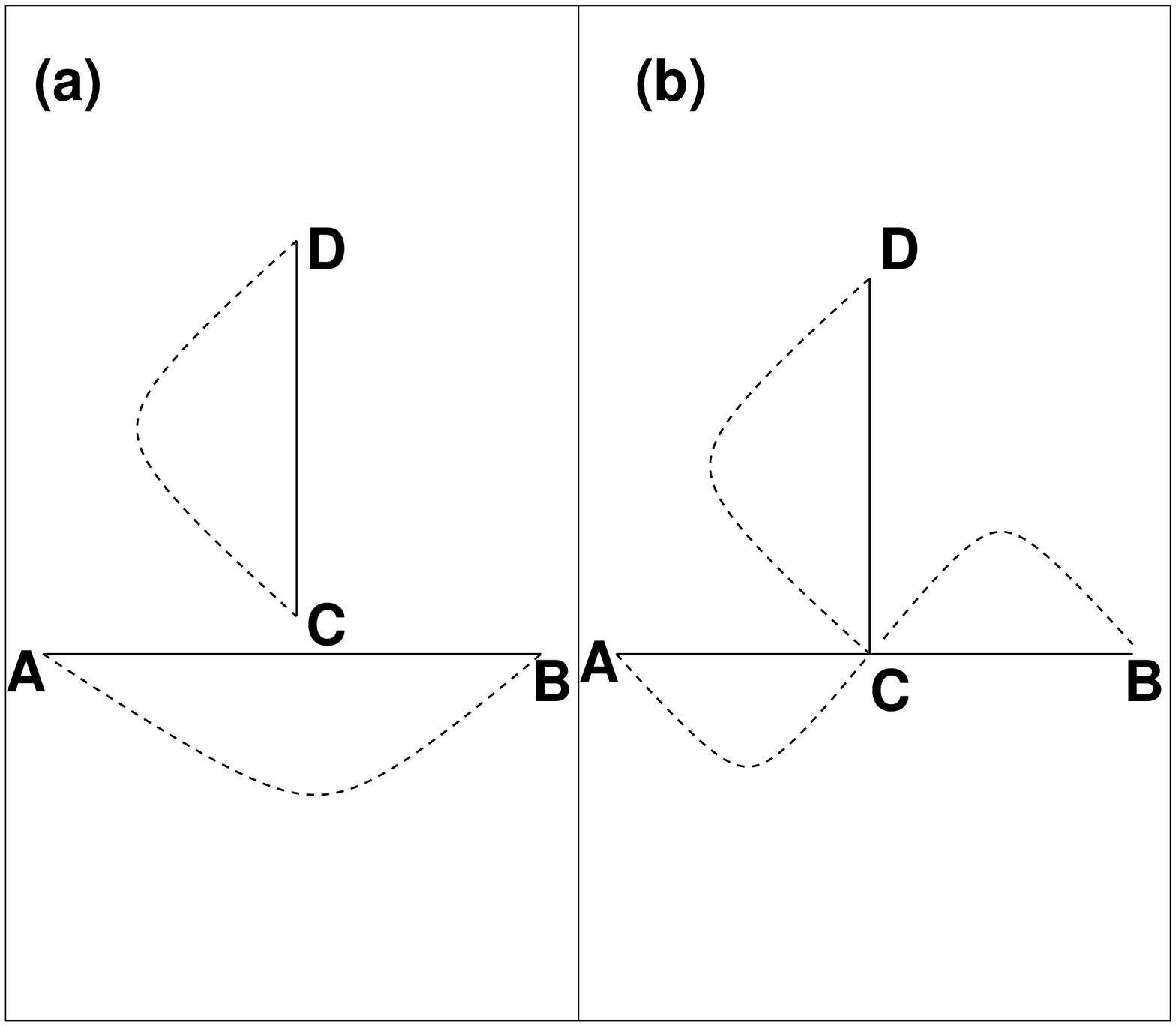}{0.5}{14}{
\hspace{-.5cm}
Two one dimensional quantum wires of equal lengths, AB and CD, shown by solid lines,
placed along x and y directions, respectively. The origin (x=0, y=0) is
at the mid-point of AB. (a) CD is not connected to AB. (b) CD is connected
to AB.
}{f2}
%%%

Note that in this case also reflection-symmetry holds in the x-direction
as well as in the y-direction, but x-y components do not separate. So by
tuning the boundary condition in y-direction by a gate voltage one can
develop nodes that try to develop across y-direction but also act across
x-direction and change the phase of the wave function in x-direction by
$\pi$.  We call them NSDN because they do not originate due to the
symmetry of the Hamiltonian. There are many configurations of such NSDN
\cite{comment}, the simplest one was discussed for the stub geometry
(shown in Fig.3) in ref.  \cite{deo96}.

Consider for example two finite one dimensional quantum wires of equal
length, AB and CD placed perpendicular to each other as shown in Fig.3 by
the solid lines.  When CD is completely detached from AB as shown in
Fig.3(a), then the quantum mechanical wave function in AB and CD in the
ground state is shown by the dotted lines. They are basically the ground
state wave function in an infinite potential well in one dimension (1D).
As is known to us, the ground state wave functions are by symmetry, even
parity states without any nodes, except at the boundary. But when CD is
attached to AB to give a T-shaped stub structure as shown in Fig.3(b),
then CD forms a node at C, which is also the midpoint of AB. The wave
function in this case is again shown by dotted lines and the wave function
between A and B is no longer an even parity state but an odd parity state.
The node at C between A and B does not originate from the symmetry of the
Hamiltonian and is not a symmetry dictated node (SDN). It is rather forced
by the boundary condition in the y-direction and is a NSDN. An
infinitesimal change in the length CD makes this node disappear and then
we have no node between A and B.  The node at C induces a phase change by
$\pi$ and when we join A and B together to form a ring-stub system, we
also get persistent currents without parity effect, as parity of the
persistent currents is sensitive to the number of nodes in the wave
function \cite{leg91}.

Fano resonances \cite{fan} are a very general feature of Q1D
\cite{bag90,tek93,deo98} systems in the presence of defects and the Fano
resonances are characterized by a zero-pole pair of the transmission
amplitude in the complex energy plane. When semi infinite leads are
attached to A and B in Fig.3(b)  then the transmission amplitude of the
stub structure also has zero-pole pair and one gets Fano resonances
\cite{tek93}. At the energy corresponding to the pole a charge gets
trapped by the scatterer and there is also an energy when there is a zero
transmission across the scatterer.  At the energy corresponding to the
zero, scattering phase shift discontinuously changes by $\pi$ due to the
NSDN \cite{deo96}.

It seems at present that this phase due to NSDN is necessary to understand
the experimental results of Refs. \cite{yac95,sch97,CER97}. Initial
analysis of the experiments in terms of Friedel-sum-rule\cite{yey95}
revealed the shortcomings of applying Friedel-sum-rule to the quantum dot.
The phase change due to NSDN can explain the experimental data was first
proposed in Ref.\cite{jay96} and later discussed in
Refs.\cite{deo98,cho98,xu98,lee99,tan99,yeycm}.

The Friedel-sum-rule (FSR) can be stated as \cite{fri52}
\begin{equation}
\th (E_{2})-\th (E_{1}) \approx \pi N(E_{2},E_{1}) .
\end{equation}
In 1D, 2D and 3D, the equality is known to be approximate,
and is almost exact in the WKB regime where generally transport
occurs.
Here \hspace{.1cm} $N(E_{2},E_{1})$ is the variation
in the number of states
in the energy interval $[E_{1},E_{2}]$ due to the
scatterer and \cite{lan61}
\be
\th  = \frac{1}{2} \sum \xi_{i} = \frac{1}{2i}
ln (det [S] ) ,
\ee
$S$ being an n$\times$n scattering matrix and $e^{i\xi_{i}}$,
i=1,2,....n are
the n eigenvalues of the unitary matrix $S$.
In differential form the FSR can also be stated as
\begin{equation}
\frac{\pa \th }{\pa E} = \frac{1}{2i} \frac{\pa}{\pa E} ln (det [S] )
\approx \pi (\rho(E) - \rho_0(E))  ,
\end{equation}

\noindent where $(\rho(E) - \rho_0(E))$ is the variation of the DOS or the
difference in the DOS due to the presence of the potential. $\rho(E)$ and
$\rho_0(E)$ can be found by integrating the local DOS $\rho(x, y, z, E)$
and $\rho_0(x, y, z, E))$ which are related to the electron probability.
However, if FSR is to be useful in mesoscopic systems where scattering
phase shifts can be accurately measured, and where we are always
interested to study a finite region of space, then we must see to what
extent it can give the local DOS. In fact in 1D, 2D or 3D it does so
extremely well and so we ask the question that how good it is in Q1D. We
shall show that in Q1D, FSR will fail to give the local DOS as well as the
global DOS at all energy regimes.

For a symmetric scatterer in a strictly 1D system when transmission and
reflection amplitudes are denoted by $t$ and $r$, respectively,
 \[ S = \left(\begin{array}{cc}
\displaystyle r & \displaystyle t \\
\displaystyle t & \displaystyle r
\end{array}\right) .\]
In this case, one can show \cite{tan99}
\be
\frac{\pa \th }{\pa E} = \frac{\pa arg(t)}{\pa E}  ,
\ee
which means sum of the phases of the eigen values of S
is equal to the phase of some particular matrix element of S.

But now we know that in systems that are not strictly 1D, one can have
zero-pole pairs and then Eq.(4) is not valid because of the $\pi$ phase
shifts induced by NSDN. Note that for the system in Fig.3(b), although the
scattering matrix is $2\times 2$, the point C is connected to 3 directions
and is hence a Q1D system. A quantum wire with a finite width and only one
propagating channel \cite{deo98} also is a Q1D system with a $2 \times 2$
scattering matrix. For such systems, that are not strictly one dimensional
but has a $2\times 2$ scattering matrix
 \bea
\frac{\pa \th }{\pa E} \ne \frac{\pa arg(t)}{\pa E} \\
\mbox{but} \hspace{1cm} \frac{\pa \th }{\pa E} \approx \pi (\rho (E)
- \rho_0 (E)) ,
 \eea
 i.e., when we go from 1D (with 2x2 S matrix)  to Q1D (also with 2x2 S
matrix) then Eq. (3) holds but Eq. (4) does not hold.  This analysis was
presented by Lee \cite{lee99} and by B{\"u}ttiker and
Taniguchi\cite{tan99}. Their analysis is restricted to the system in
Fig.3(b) and $S$-matrices that are 2$\times$2.  Eq.(3) is not violated in
the presence of NSDN and $\pi$ phase slips because $det[S]=r^2-t^2$ and if
$arg(t)$ changes by $\pi$ then $[arg(t^2)]$ changes by $2\pi$ or $0$ and
hence $det[S]$ is unaffected by the $\pi$ phase slips. Thus for such a
single propagating channel this phase shift is well understood by now.  It
has been emphasized that the multi-channel case also needs to be studied
\cite{yeycm}, specially since scattering phase shifts can now be probed
experimentally in the single channel case \cite{yac95,sch97} as well as in
the multi-channel case \cite{CER97}, but no such study has been reported
so far.

In particular, the FSR is very important in condensed matter Physics,
because from the scattering phase shift (that can be determined
experimentally) one can know the local DOS inside a disordered sample
without knowing its internal details.  Even theoretically, except in very
simple situations, the wave function inside the scatterer has infinite
degrees of freedom and exact calculation of local DOS from the exact wave
function inside the scatterer may be non-trivial. While a good estimate of
local DOS from the S-matrix, which is on its own a very useful quantity,
can greatly reduce the complexities.  In this work we will study the $n$
channel scattering problem in a Q1D quantum wire, with special emphasis on
FSR and Wigner delay time \cite{smi59,kumar}.  Refs. \cite{lee99} and
\cite{tan99} parameterize the $S$-matrix in a particular way (there are in
fact many different ways of parameterizing the $S$-matrix ) in which the
scattering matrix elements become independent of energy. We will show that
this energy dependence, that are not important in 1D play a very crucial
role in Q1D multi-channel scattering. Hence in section II we will
generalize the work of Refs.\cite{lee99} and \cite{tan99} for real energy
dependent 2$\times$2 scattering matrices. The $n$ channel case will be
analyzed in section III and IV. In section V we will show some novel phase
shifts at critical energies where $S$ matrix changes dimensions. Section
VI is devoted to conclusions.

%%%%%%%%%%%%%%%%%%%%%%%%%%%%%%%%%%%%%
\section{Scattering in one-dimension and negative values of $d \th /dE$}
\label{s2}
%%%%%%%%%%%%%%%%%%%%%%%%%%%%%%%%%%%%%

In Fig.4 we consider a potential that is described in details in the
figure caption. The quantum mechanical wave function or the solution to
the Schr{\"o}dinger equation in different regions is also shown and
explained in the figure and its caption. We will always normalize the
incoming wave-function such that its amplitude is 1. Griffiths boundary
conditions for this system gives the following equations \cite{gri,deo93}
(we use $2m=1$ and $\hbar=1$).
\begin{equation}
1 + r = a + b ,
\end{equation}
\begin{equation}
a e^{ikl} + b e^{-ikl} = t ,
\end{equation}
\begin{equation}
ik(1-r) - ik(a-b) = - V (1 + r)\hspace{.5cm}\mbox{and}
\end{equation}
\begin{equation}
ik(a e^{ikl} - b e^{-ikl}) -ikt = - V(a e^{ikl} + b
e^{-ikl}) .
\end{equation}
%%%
\PostScript{6}{0.5}{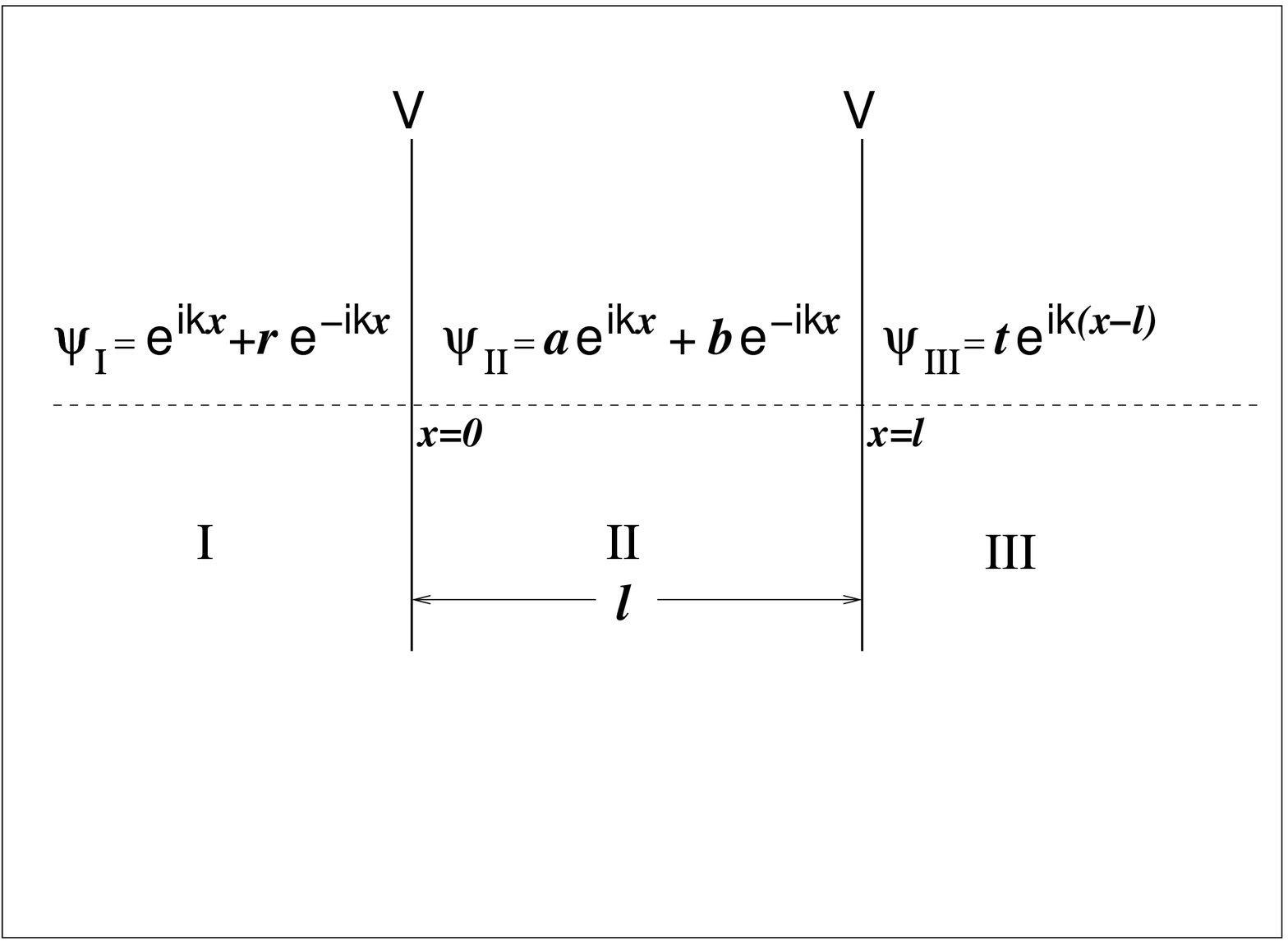}{0.5}{14}{
\hspace{-.5cm}
  Two identical delta function potentials separated by a length $l$.
Strength of each potential is $V$. The thick vertical lines denote the
positions of the potentials and the thin horizontal line is the direction
of propagation.  A plane wave of unit amplitude is incident from the left
and wave function in different regions (marked as I, II and III) is
written down in the figure. $r$ and $t$ are the reflection and
transmission amplitudes, respectively, of the entire system and $ k=\sqrt
{E} $ is the incident wave vector.
The origin of coordinates is shown in the figure.
}{f1}
%%%
 \noindent We will first analyze this system in detail and generalize the
results of Refs. \cite{lee99,tan99} further by considering realistic
energy dependent $r$ and $t$, that will later help us to accentuate the
new features that can be observed in a multi-channel disordered quantum
wire.

First of all let us calculate the local DOS and
global DOS to see how much it agrees
with $d \theta /dE$. Although the basic facts
discussed in this section is known in the Greens
function formalism, to the best
of our knowledge, quantitative disagreement (or agreement)
has not been shown so far.  
Using
quantum mechanical expression for the local DOS integrated over the region
II in Fig.~4, i.e.,
 $$ \rho_R = {2 \over hv} \int ^l_0 \mid a e^{ikx} + b
e^{-ikx} \mid ^2 dx, $$
 it is easy to show from Eq. 3 that (the Eq. below
is consistent with Ref. \cite{tan99} and some hints on its derivation
is given in Ref \cite{com})
 \be
{d \theta \over d (kl)} \approx {
\bar \rho } + {\rho _q \over l} = \rho '\hspace{.5cm} \mbox{(say)},
\ee
\be
\mbox{where}\hspace{.5cm} { \bar \rho } = \mid a \mid ^{2} + \mid b
\mid ^{2} \hspace{.1cm}\mbox{and}
\ee
\be \rho_{q} = \int ^{l}_{0}
\left(ab^* e^{2ikx} +ba^* e^{-2ikx} \right) dx .
\ee
 Here $\frac{\rho_{q}}{l}$ is a term that arises because of quantum
mechanical interference and it can be seen that the integrand in Eq.(13)
oscillates with $x$. For $|\frac{E}{V}|>1$ (this is the WKB regime
when the electron does not feel the potential strongly and is
almost entirely transmitted) $\frac{\rho_{q}} {l}$ is
negligibly small. This is shown in Fig.5, where we plot $\rho^{'}$ (the
dashed curve) and ${\bar \rho }$ (the dotted curve). The two curves are
almost the same for $|{E \over V}|>1$, which means ${\rho_q \over l}$,
being the difference between the dashed and dotted curves is vanishingly
small above this energy.  It is known that to get the equality between the
LHS and RHS of Eq. 11, it is necessary to drop the term ${\rho_q \over
l}$ \cite{gas}.  
It is also known that this deviation arises because we are
considering the local DOS rather than the global DOS.
If we consider the global DOS, i.e.,
$$\rho(E)={2 \over hv} \int^{\infty}_{-\infty} \psi^*(x) \psi(x) dx$$
where $\psi(x)$ is the quantum mechanical wavefunction at x,
then instead of Eq. (11) we get
$${d \theta \over d(kl)} \approx
Lt_{l \rightarrow \infty} (|a|^2 + |b|^2).$$
The equality is still approximate although almost exact in the WKB
regime because the RHS is positive definite, while it is well known
that the LHS can become negative at low energy (non-WKB regime)
as will be demonstrated below.
To get the equality it is
necessary to neglect something else as we shall soon see.  This second thing
that we shall drop is however not due to the fact that we are considering
the local DOS. It is an inherent approximation of the FSR even when we are
considering the global DOS and hence also when we are considering the
local DOS.  The $\rho_q/l$ term or the interference term inside the
scatterer does not arise in the case when $l \rightarrow 0$ as in the case
to be considered in section IV.  All the deviation to be observed there is
due to this inherent weakness of the FSR. We shall also see below that
this inherent weakness of the FSR is negligible in 1D, 2D or 3D but
becomes very formidable in quasi 1D.

%%%
\PostScript{6}{0}{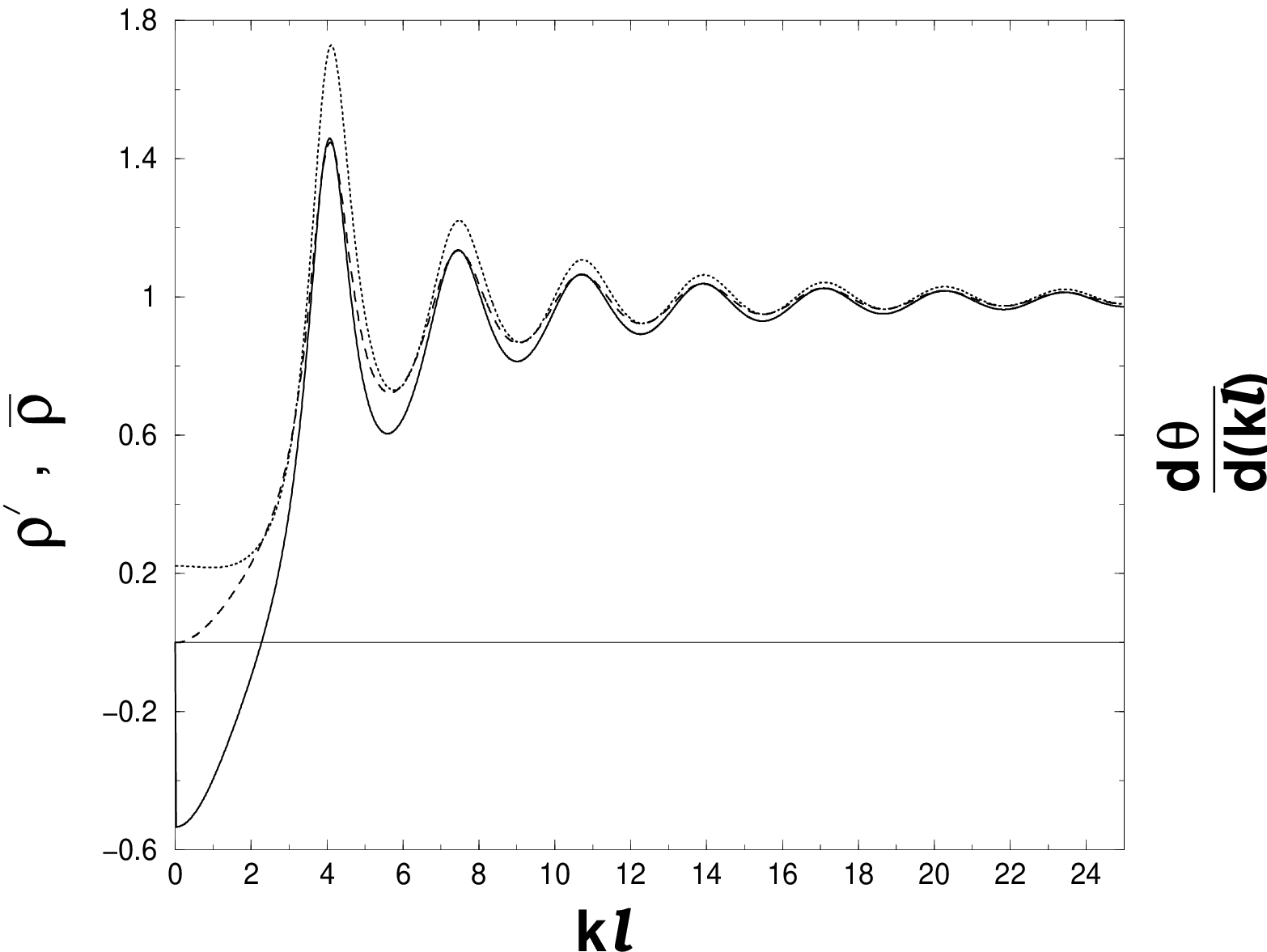}{0.5}{14}{
\hspace{-.75cm}
 The system under consideration is shown in Fig.4. The solid curve gives
the exact $d \theta/d(kl)$, the dashed curve gives the $\rho^{'}$ and the
dotted curve gives ${\bar \rho}$.  This plot is done for $Vl^2=-5,\hbar=1,
2m=1$.
 }{f2}
%%%

%%%
\PostScript{6}{0}{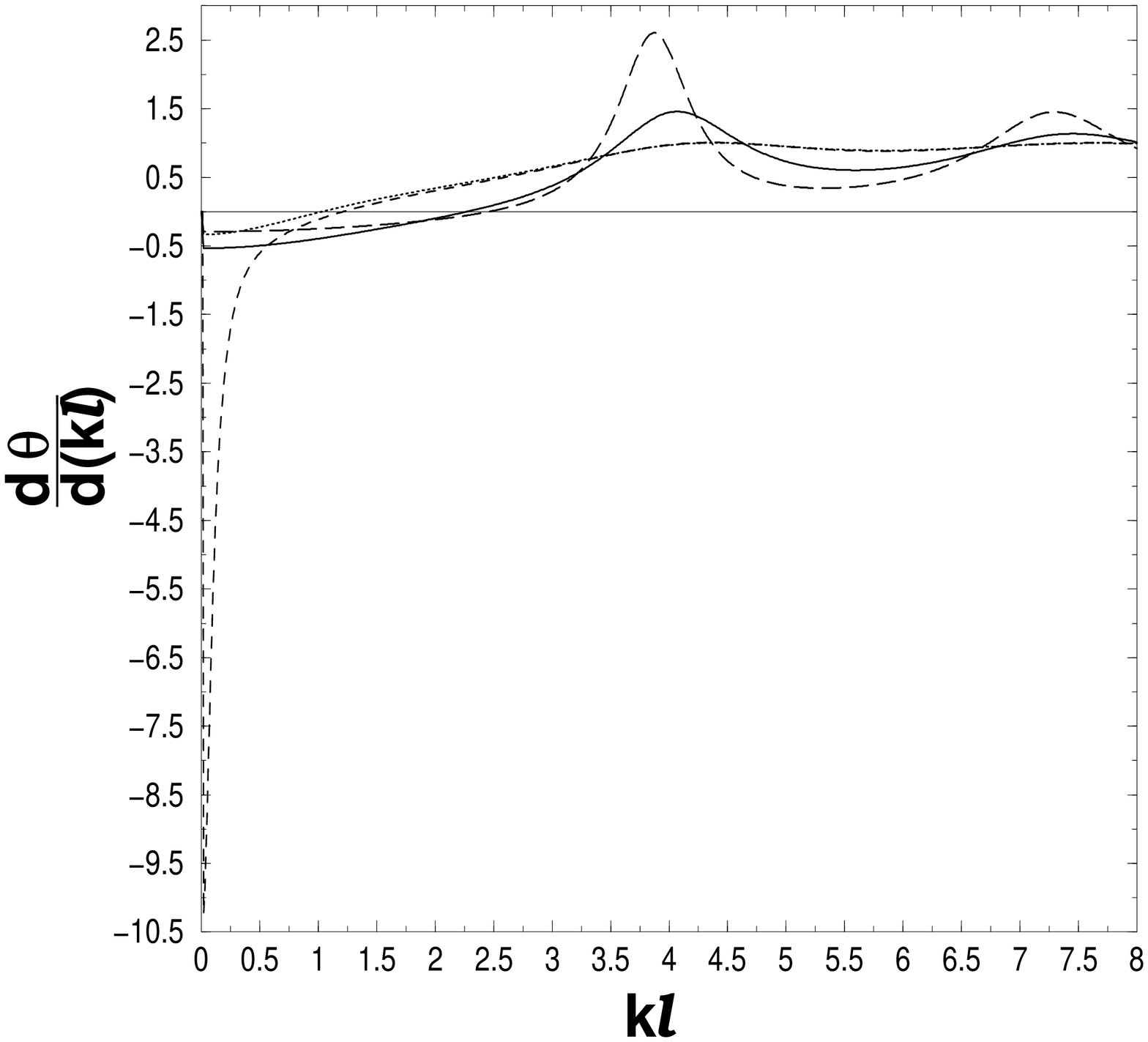}{0.5}{14}{
\hspace{-.5cm}
 The system under consideration is shown in Fig.4.  The plot is of $d
\theta/d(kl)$ versus $kl$ for the system for different values of $Vl^2$.
The dotted curve is for $Vl^2=-2 $, the dashed curve is for $Vl^2=-2.1$,
the solid curve is for $Vl^2=-5$, the long dashed curve for $Vl^2=-8$. We
use $\hbar=1, 2m=1$.
 }{f2}
%%%
 One can prove that \[{ \bar \rho } = \mid a \mid^{2} + \mid b \mid^{2} =
\frac{1-\mid r^{'}\mid^{4}}{\mid 1-{r^{'}}^{2}\tau^{2}\mid^{2}} ,\] for
any energy dependent reflection amplitude $r^{'}$ of one of the two
identical scatterers in Fig.4, where, $\tau=e^{ikl}$.
Hence as indicated by Eq. 11, it would be
interesting if we can obtain a good estimate of ${\bar \rho}$ or $\rho'$
from $\th$.  In the appendix I it is explained that if $\frac{d
r'}{dE}=\frac{d t'}{dE} \rightarrow 0$ (which means the
scatterers are non-dispersive and that
only happens at high
energy in 1D, 2D and 3D)  then, $\frac{d \theta}{d (kl)}$ reduces to the
expression $\frac{1-\mid r^{'}\mid^{4}}{\mid
1-{r^{'}}^{2}\tau^{2}\mid^{2}}$, and then therefore, ${d \th \over d(kl)}
= {\bar \rho}$. It is shown in section III of Ref. \cite{yeycm} (see Eq. 6
and 7 therein), that to relate ${d \theta \over dE}$ to the global DOS, one
has to neglect the energy dependence of the self energy, that depends on
the coupling of the system to the leads, i.e., $r'$ and $t'$. Thus our
results are consistent with that. The exact
$\frac{d\th}{d(kl)}$ is shown in Fig.5 by the solid curve.  Note that in
the relevant energy regime ($|\frac{E}{V}|>1$),
the solid curve is very close to the dashed and dotted curves,
which means FSR works very well for the local DOS
as well as for the global DOS. But for
$|\frac{E}{V}|<1$, $\frac{d\th}{d(kl)}$ deviates from ${\bar \rho}$.  But
since transport effects in weak localization or diffusive or ballistic
regime occur at Fermi energies, that is normally higher in semiconductors
as well as metals in comparison to the energy where the two curves deviate
substantially from each other, Friedel sum rule is often useful in
condensed matter to obtain a good estimate of local DOS as well as global
DOS.

$d \th / dE$ is also well known as Wigner delay time \cite{smi59,kumar}.
In the stationary phase approximation, it gives the time spent by the
scattered particle at the impurity site.  In the low energy regime, where
dispersion becomes significant, the stationary phase approximation is
not valid and $d \th /dE$ can become negative and does not give a
meaningful particle
delay time.  In this regime $d \th / dE$ becomes negative as
the phase velocity becomes larger than the group velocity and even larger
than the velocity of light, and although such super-luminous particles can
be detected experimentally they cannot carry any signal or information.  In
Fig.6 we show the negative behavior of $d \theta/d(kl)$.  We find that as
the strength of the impurities is varied, $d \theta /d(kl)$ can become
more or less negative (see Fig.6), maximizing at $Vl^2=-2.1$ for the
symmetric delta potentials. The energy
regime, where $d \theta /d(kl)$ can be negative remains the same for all
$V$ and always $|\frac{E}{V}|<1$. We have checked for all these values of
$V$ that apart from this insignificant energy range, FSR works very well.
FSR has a close counterpart in quantum mechanics called Levinson's
theorem.  It is known that Levinson's theorem also breaks down in the
presence of zero energy bound states \cite{new82} that can be degenerate
with scattering states.

%%%%%%%%%%%%%%%%%%%%%%%%%%%%%%%%%%%%%
\section{ Wigner delay time in quasi-one-dimensions}
\label{s3}
%%%%%%%%%%%%%%%%%%%%%%%%%%%%%%%%%%%%%

%%%
\PostScript{6}{0}{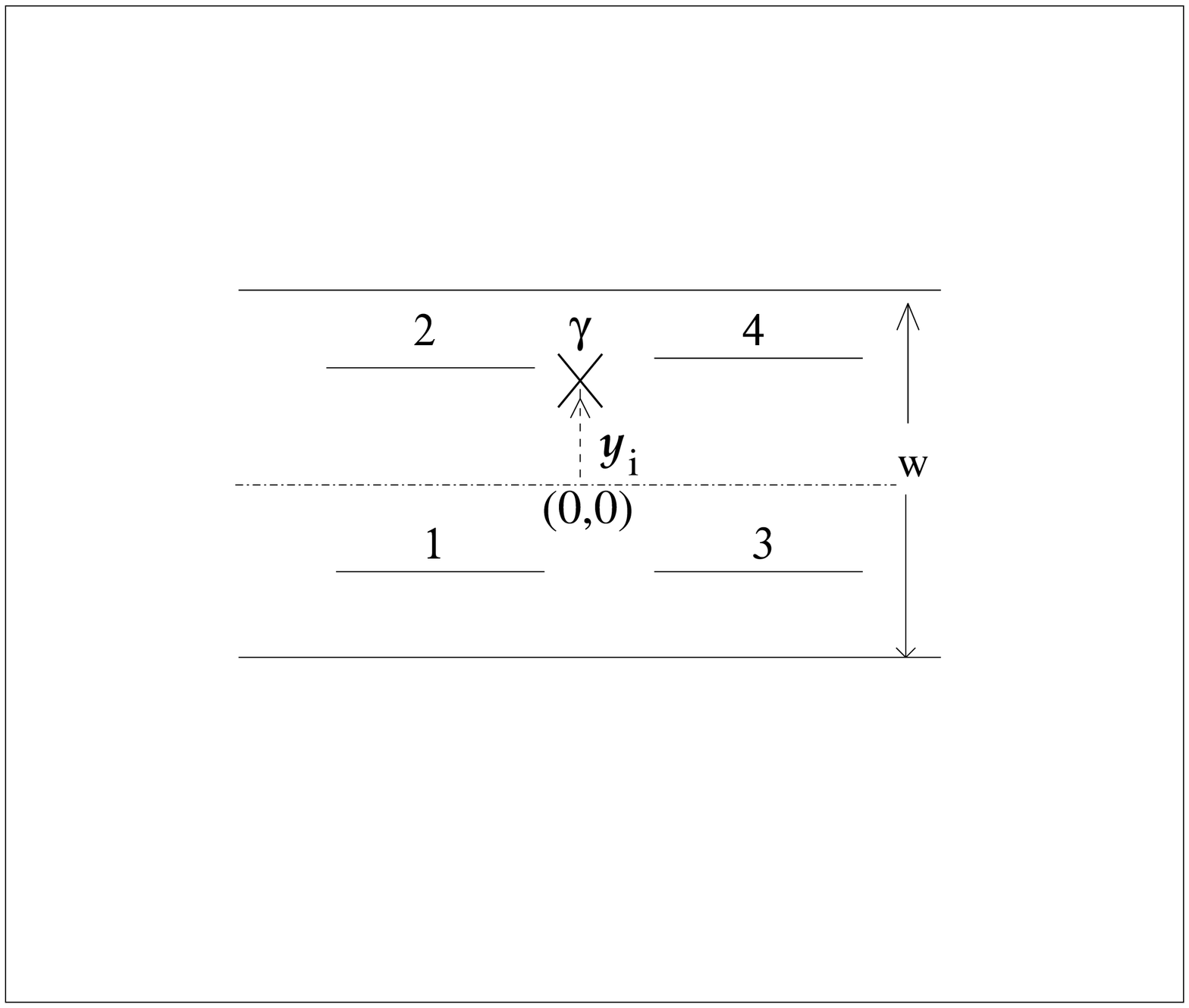}{0.5}{14}{
\hspace{-.5cm}
 Here we show a quantum wire of width $W$. The dash-dotted curve is a line
through the middle of the quantum wire, and it is also taken to be the
x-axis. The origin of the coordinates is shown in the figure. A delta
function potential $V(x,y)=\gamma \delta(x)  \delta(y-y_{i})$ is situated
at $x=0$ and $y=y_{i}$ and marked as $\times$. We consider scattering
effects when the incident electron is from the left. The sub-bands on the
left of the impurity is denoted as 1 for the first mode (i.e., its wave
function can be obtained by putting $n$=1 in Eq.(14) with appropriate sign
for $k_n$) and 2 for the second mode (i.e., its wave function can be
obtained by putting $n$=2 in Eq.(14)  with appropriate sign for $k_n$).
Similarly the sub-bands on the right of the impurity is denoted as 3 for
the first mode (i.e., its wave function can be obtained by putting $n$=1
in Eq.(14) with appropriate sign for $k_n$)  and 4 for the second mode
(i.e., its wave function can be obtained by putting $n$=2 in Eq.(14) with
appropriate sign for $k_n$).  The impurity at $\times$ mixes these wave
functions to give a scattering matrix element $t'_{mn}$ from mode $m$ to
mode $n$.
 }{f2}
%%%

In Fig.7 we consider a quasi-one-dimensional quantum wire with an
attractive impurity at (0, $y_i$), having electrons confined along the
y-direction but free to move along the x-direction.  While the states far
away from the impurity are good momentum states, the impurity can mix the
different modes and in this region of mode mixing, the wave function is
$\psi (x,y)= \Sigma_n c_n(x) \chi_n(y)$, where $\chi_n(y)$ are the
transverse wave functions in the absence of the impurity and $c_n(x)$ are
position dependent coefficients that has to be determined by mode
matching. The confining potential in the y-direction or the transverse
direction is taken to be hard wall. Thus the transverse wave-function is
of the form $\chi_n(y)=Sin \frac{n\pi}{W}(y+\frac{W}{2})$.  For a given
width $W$ of the quantum wire one can choose the energy range of the
incident electron such that only two modes are propagating, although, all
the other modes (infinite in number, showing that the internal wave
function can have infinite degrees of freedom, which makes it difficult to
calculate the exact local
DOS from the internal wave function) will be present
but as evanescent modes.  For example, if the energy of the electron be
$E$ then for propagation in the n-th transverse mode (in short we will
refer this as n-th mode) the wave-function is of the form
\be
Sin \frac{n\pi}{W}(y+\frac{W}{2})\hspace{.2cm}e^{ik_{n}x}
\ee
where  $k_{n}= \sqrt{E-E_n}$, $E_n$ being
$\frac{n^2\pi^2}{W^2}$ \hspace{.2cm} and \hspace{.2cm} $n=1,2,3,
\cdots$ $\infty $. Here we have used $\hbar=2m=1$.
To have the n-th mode to be propagating it is necessary that
$k_{n}^2 > 0 $ or
\be
n < \frac{W}{\pi} \sqrt{E}.
\ee Thus we can choose the energy range where there
will be two propagating modes, i.e., $n=1$ and $n=2$ satisfy condition (15).
The rest of the modes ($n>\frac{W}{\pi}\sqrt{E}$) will be evanescent, whose
wave functions are of the form
\be
Sin \frac{n\pi}{W}(y+\frac{W}{2})\hspace{.2cm}e^{- \kappa_{n}x} ,
\ee
where $\kappa_n=\sqrt{E_n-E}$. These evanescent modes just renormalize the
scattering matrix elements and drop out of the problem.
The transmission amplitude from m-th incident mode to n-th
scattered mode is given by \cite{bag90}
\bea
t_{mn}^{'}  =
-\frac{i\Gamma_{mn}}{2d\sqrt{k_{m}k_{n}}},\\
\mbox{where} \hspace{.5cm} d = 1+\sum ^{e}\frac{\Gamma_{nn}}{2\kappa_{n}}+
i\sum ^{p}\frac{\Gamma_{nn}}{2k_{n}}.
\eea
Here $\sum ^{e}$ denotes the sum over all evanescent modes and $\sum ^{p}$
denotes the sum over all propagating modes. Eq.(17)  holds only for
inter-subband transmission amplitudes and all reflection amplitudes.  When
we say reflection amplitude we mean the following. For an electron
incident from the left, all outgoing channels to the left are reflection
channels.  According to this convention $t_{11}^{'}$, $t_{22}^{'}$,
$t_{21}^{'}$ and $t_{12}^{'}$ are reflection amplitudes. Inter-subband
transmission amplitude are then obviously $t_{14}^{'}$ and $t_{23}^{'}$.
The intra-subband transmission amplitudes $t_{13}^{'}$ and $t_{24}^{'}$
(according to numbering of channels explained in Fig.7) are given by
\be
t_{13}^{'} = 1+t_{11}^{'}
 \hspace{.5cm}\mbox{and}\hspace{.5cm}
 t_{24}^{'} = 1+t_{22}^{'}.
\ee
Here $ \Gamma_{nm} $ is the strength of coupling between the n-th mode and
the m-th mode. If we take the impurity to be a delta function potential
i.e., $V(x,y)= \gamma \delta(x) \delta(y-y_i)$, and the confining
potential in the y-direction to be hard wall ($V=\infty$ for
$-\frac{W}{2}\ge y\ge \frac{W}{2}$, and $0$ everywhere else except the
impurity site $\times$ ) (see Fig.7) then
\[  \Gamma_{nm} = \gamma Sin \frac{n \pi}{W}(y_{i}+\frac{W}
{2})
Sin \frac{m \pi}{W}(y_{i}+\frac{W}{2}). \]

Apart from the two propagating modes we consider two evanescent modes and
truncate the infinite series of evanescent modes (note that although the
series is strongly converging, the reason for truncating is different,
stronger and explained in more detail after Fig. 12) in Eq. (18) and so
Eq. (18) becomes \\
\be
1+ \frac{\Gamma_{33}}{2\kappa_{3}}+\frac{\Gamma_{44}}{2\kappa_{4}}
+i\left( \frac{\Gamma_{11}}{2k_{1}}+\frac{\Gamma_{22}}{2k_{2}}\right)=d_2
\hspace{.2cm} \mbox{(say)}.
\ee
The lowest evanescent mode (putting $n$=3 in Eq.  (16) gives its wave
function)  has even parity in the transverse direction. For a negative
impurity potential $i.e.$, $\gamma < 0$, it also has a bound state at
$E=E_{3b}$, where $E_{3b}$ is given by the solution of
\be
1+\frac{\Gamma_{33}}{2\kappa_{3}}+\frac{\Gamma_{44}}{2\kappa_{4}}=0.
\ee
Since $E_{3b}<{9 \pi^2 \over W^2}$, $E_{3b}$ can be degenerate with
scattering states (the $n$=1 and $n$=2 modes are the scattering states).
The higher evanescent mode (putting $n$=4 in Eq. (16) gives its wave
function) has odd parity in the transverse direction and this too has a
bound state at $E=E_{4b}$, where $E_{4b}$ is given by the solution of \be
1+ \frac{\Gamma_{44}}{2\kappa_{4}}=0.  \ee Once again depending on
$\gamma$, $E_{4b}$ can be degenerate with the scattering states.  The
effect of including more evanescent modes is just to renormalize the
strength of the impurity potential and does not give anything new
\cite{bag90}.

The scattering matrix in this case is
\be
S = \left[ \begin{array}{llcl}
t_{11}^{'} & t_{12}^{'} & t_{13}^{'} & t_{14}^{'} \\
t_{21}^{'} & t_{22}^{'} & t_{23}^{'} & t_{24}^{'} \\
t_{31}^{'} & t_{32}^{'} & t_{33}^{'} & t_{34}^{'} \\
t_{41}^{'} & t_{42}^{'} & t_{43}^{'} & t_{44}^{'}
\end{array} \right]
=
\left[ \begin{array}{cc}
r_{2c} & t_{2c} \\
\hat{t}_{2c} & \hat{r}_{2c}
\end{array} \right] ,
\ee
\[ \mbox{where}\hspace{.5cm} r_{2c} =
\left[ \begin{array}{cc}
t_{11}^{'} & t_{12}^{'} \\
t_{21}^{'} & t_{22}^{'}
\end{array} \right]
\]
\[\mbox{and} \hspace{.5cm} t_{2c} =
\left[ \begin{array}{cc}
t_{13}^{'} & t_{14}^{'} \\
t_{23}^{'} & t_{24}^{'}
\end{array} \right] .
\]
$\hat{t}_{2c} = t_{2c}$ due to time reversal symmetry and $\hat{r}_{2c} =
r_{2c}$ for a symmetric scatterer as that considered here.  Now once again
due to micro-reversebility $t_{12}^{'} = t_{21}^{'}$. Also $t_{12}^{'} =
t_{14}^{'}$ because in both $t_{12}^{'}$ and $t_{14}^{'}$ the density of
states in the input as well as the output channel is the same, and also
the incident channel momenta and the outgoing channel momenta are the same
in the transverse as well as in the propagating direction.  Also
$t_{23}^{'} = t_{41}^{'}$ because transmission amplitude should be
independent of the position of the observer i.e., whether the observer is
looking into the plane of the paper or out of the plane of the paper. Thus
among the 16 matrix elements in Eq. (23) we are left with only 5 that are
distinct. They are $t_{11}^{'},t_{12}^{'},t_{22}^{'},t_{13}^{'}$ and
$t_{24}^{'}$.\\ From Eq. (17) and (19),
\bea
t_{11}^{'}  = -\frac{i\Gamma_{11}}{2d_2k_{1}},\\
 t_{12}^{'}  = -\frac{i\Gamma_{12}}{2d_2\sqrt{k_{1}k_{2}}},\\
 t_{22}^{'}  = -\frac{i\Gamma_{22}}{2d_2k_{2}}, \\
 t_{13}^{'}  = \frac{1+ \frac{\Gamma_{33}}{2\kappa_{3}}+\frac{\Gamma_{44}}
{2\kappa_{4}}+i\frac{\Gamma_{22}}{2k_{2}}}
{d_2} \hspace{.5cm}\mbox{and}\\
t_{24}^{'}  = \frac{1+ \frac{\Gamma_{33}}{2\kappa_{3}}+\frac{\Gamma_{44}}
{2\kappa_{4}}+i\frac{\Gamma_{11}}{2k_{1}}}
{d_2}\hspace{.2cm}.
\eea
Knowing these matrix elements, the scattering matrix is completely known
and $\th $ can also be calculated.

We find some further relationships between the scattering phase shifts as
follows. First of all
\be
arg(t_{11}^{'}) = arg(t_{22}^{'}) =tan^{-1}\frac{Re(d)}{Im(d)}  .
\ee
Secondly, when $\frac{4\pi^2}{W^2} <E_{3b}<\frac{9\pi^2}{W^2}$, i.e., the
bound state of the 3rd subband lies in the energy range where one can have
two propagating subbands, then the bound state $E_{3b}$ drastically
changes the scattering matrix elements in that energy range.  So in this
energy range $\frac{4\pi^2}{W^2}$ to $\frac{9\pi^2}{W^2}$ we find
\be
arg(t_{12}^{'}) \mp \frac{\pi}{2} = \th + \pi  .
\ee
Here negative sign is to be taken when $E_{3b}$ lies in this energy range.
Otherwise the positive sign has to be taken. $\th$ is to be calculated
from Eq.  (2) using Eq. (23). Thirdly we find
 \be
arg(t_{11}^{'}) \pm \pi = arg(t_{12}^{'})  .
 \ee
 Note that
in contrast to Eq. (30) here the
choice of $\pm$
sign is arbitrary. However consistent with this choice is the
following
 \be
arg(t_{11}^{'}) \pm \frac{\pi}{2} = \th + \pi ,
 \ee
 where once again + sign is to be
taken when $E_{3b}$ is present in this energy
range and - sign is to be taken when absent.

We thus find very simple analytical expressions for $\th$ in the sense
that one need not calculate it from a $4\times4$ scattering matrix but can
calculate it from the argument of a single matrix element like
$t_{11}^{'}$ or $t_{12}^{'}$ or $t_{22}^{'}$. These relations are
analogous to Eq. (4) in section I obtained for purely one dimensional
case, i.e., one need not calculate $\th$ from 2$\times$2 matrix but one
can find it from the argument of a single matrix element.

%%%
\PostScript{6}{0}{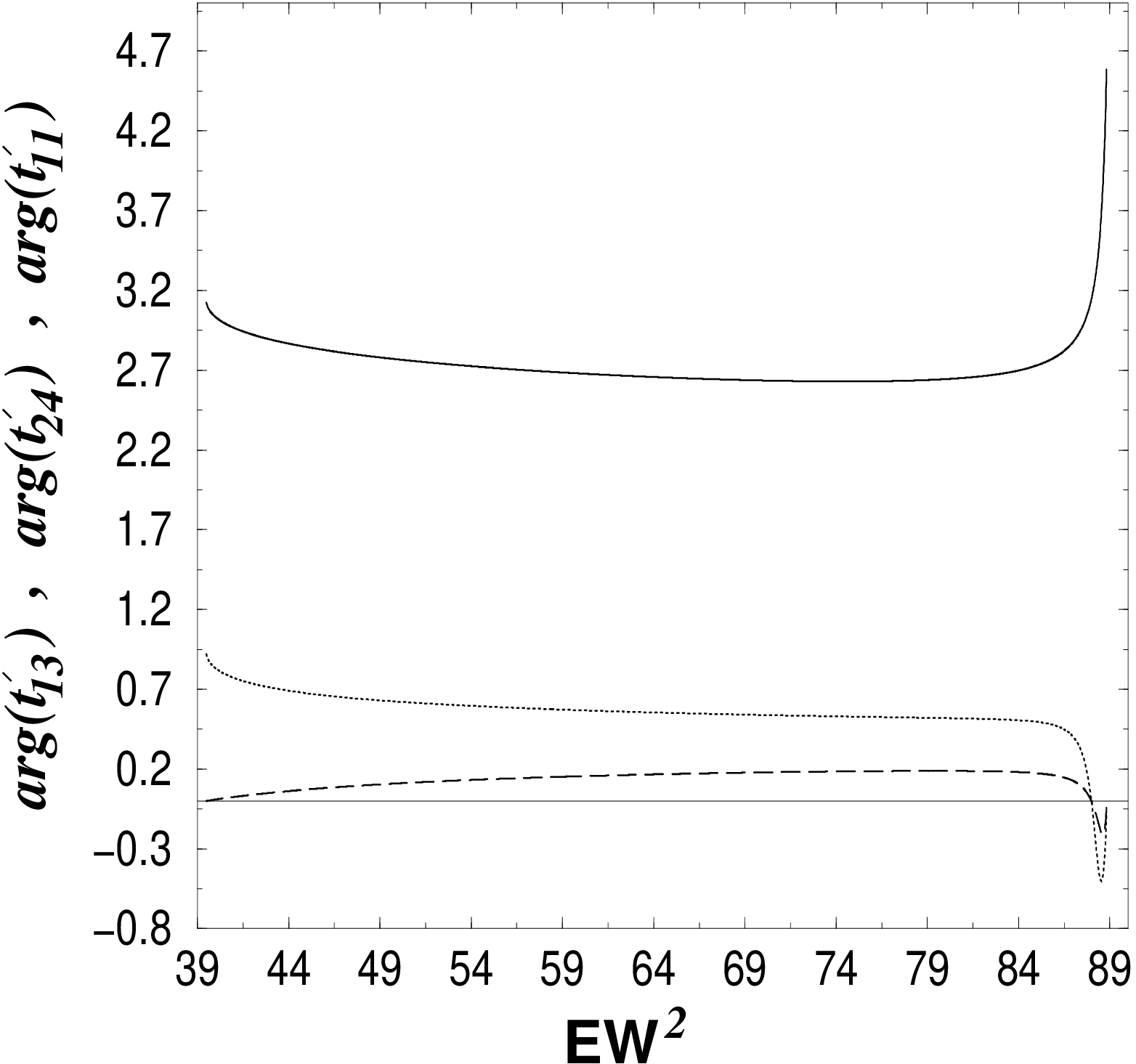}{0.5}{14}{
\hspace{-.5cm} The system under consideration is shown
in Fig. 7. The plot is of the argument of various transmission amplitudes
($arg(t_{mn}^{'})$ in radians)
from incident channel $m$ to propagating channel $n$ versus
E$W^2$ . The solid curve gives $arg(t_{11}^{'})$,
the long dashed curve gives $arg(t_{13}^{'})$ and
the dotted curve gives $arg(t_{24}^{'})$.
We use $\gamma=-10$, $y_{i}=.21 W$
and $x_i = 0$
}{f2}
%%%

In Fig.~8, we plot only the distinct arguments of the scattering
amplitudes versus energy of the incident electron. We find that all of
them show negative slopes over a very large range of energy and as already
discussed, such negative slopes give rise to fundamental questions in
quantum mechanics \cite{smi59,kumar}.  Now in Q1D we find that
this negative slope is not restricted to low energy but can occur at any
arbitrary energy.  Notice for example, $arg(t'_{13})$ and $arg(t'_{24})$
show larger negative slopes at the highest possible energies for two
channel propagation.  The rest of this section will be devoted to
understanding these negative slopes that at first sight looks very
different in nature and character in the three curves in Fig.~8,
and also to understanding what will happen when there are more
than two propagating modes. We will
address the FSR in the next section.

It is to be noted that among all these scattering matrix elements
$t_{11}^{'}$ and $t_{13}^{'}$ exist in the single channel regime
(i.e., $ \pi^2< EW^2 < 4\pi^2$) where
$t_{11}^{'} $ is the reflection amplitude and $t_{13}^{'} $ is the
transmission amplitude. The phase of $t_{13}'$ in the single channel
regime is known to change discontinuously by
$\pi$ when $t_{13}'$ is 0, i.e., $t_{13}'$ has a zero
in real energy. In the two channel
regime if we write from simplifying Eq. (27)
\be
t_{13}^{'} =\frac{k_2(2\kappa_3+g_3)+i\kappa_3
g_2}{k_2(2\kappa_3+g_3)+i\kappa_3(g_2+\alpha
g_1)},
\ee
where $\alpha=\frac{k_2}{k_1}$ and
$ g_s = \frac{2\kappa_4}{\Gamma_{44}+2\kappa_4}\hspace{.2cm}
\Gamma_{ss}$; with $s$=1,2,3.
then interestingly, we see that it has a zero
in complex energy and not in real energy.

If we modify the Breit-Wigner line shape formula of 1D to
include complex zeroes and write
\be
t_{mbw}(E) = A\hspace{.1cm}\frac{E-E_0+i\Gamma _0}{E-E_p+i\Gamma _p}
\hspace{.1cm},
\ee
where A is a
normalization factor, then just as $\Gamma_p$ gives the scale over
which $arg[t_{mbw}(E)]$ increase at $E=E_p$, $\Gamma _0$ gives a scale
over which $arg[t_{mbw}(E)]$
decrease at $E=E_0$ where $|t_{mbw}(E)|^2$ also shows a minimum
at $E=E_0$ (but not zero).
One can check this very easily (let us say, when
$E_0$=2, $E_p$=1 and $\Gamma_0=\Gamma_p=0.5$) and so we do not
demonstrate it here. Now from Eq. (33) we
see that at an energy which satisfies the condition
\be
2\kappa_3+g_3=0 ,
\ee
the real part of the numerator in Eq. (33) is zero. Condition (35)
is the same as the condition (21) for a
bound state $E_{3b}$ coming from the 3rd subband
that is degenerate with scattering states.
So, around this energy where Eq.(35) is satisfied (lets say at $E=
E_{3b} \equiv E_0$)
$arg(t_{13}')$ will undergo a drop over an energy scale
determined by the imaginary part, $\kappa_3 g_2 $, i.e.,
$\Gamma _0\equiv\kappa_3 g_2$.

It can be seen in Fig.9 that $|t_{13}^{'}|^2$ (dotted curve)  shows a
narrow minimum around an energy E$W^2\simeq 84$ (which is the solution of
Eq. (35) or Eq. (21))  and at this energy $arg(t_{13}^{'})$ shows a very
sharp drop over a narrow energy range determined by $\kappa_3 g_2$.  Hence
by decreasing/increasing this quantity $\kappa_3 g_2$ we can make the
phase drop sharper/broader.  $g_2$ can be made smaller in two ways, first
by decreasing $\gamma$ and second by taking the impurity closer to a node
in the transverse wave function. The plot for a decreased value of
$\gamma$ is shown in Fig.10 and it confirms this.

%%%
\PostScript{6}{0}{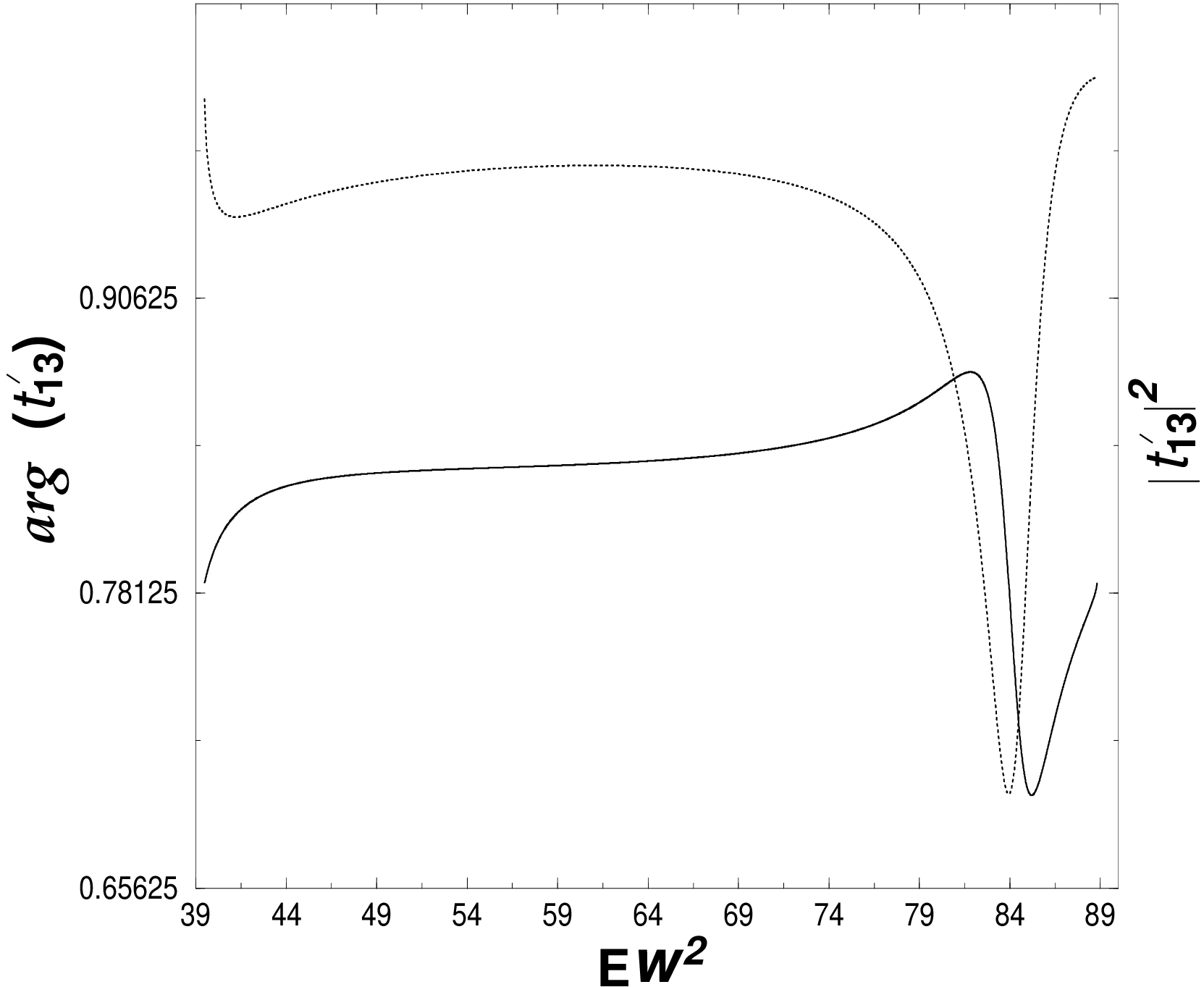}{0.5}{14}{
\hspace{-.5cm}
The system under consideration is shown in Fig.7.
The solid curve gives $arg(t_{13}^{'})$ in radians shifted by
$\frac{\pi}{4}$ radians in the
y-direction and the dotted curve gives $|t_{13}^{'}|^2$.
Both the functions are plotted versus
$EW^2$ using $x_i=0$, $y_i=.45W$ and $\gamma=-15$.
}{f2}
%%%

Note that the quantity $\kappa_3 g_2$ is actually energy dependent.  But
in Fig.9 and Fig.10 $\kappa_3 g_2$ is so small that the drop occurs over a
scale in which $\kappa_3 g_2$ is roughly constant. For larger values of
$\kappa_3 g_2$, the phase drop will be determined by a complex competition
between $\kappa_3$ and $g_2$. This is shown in Fig.11.  First of all the
scale of the phase drop becomes so large that any sensitivity to the
position of the bound state can not be seen. Secondly, $\kappa_3 g_2$ can
not be taken to be a constant over this large scale and the enhancement of
the negative slope for $EW^2 > 79$ is a signature of the fact that here
$\kappa_3 \rightarrow 0$ and so $\kappa_3 g_2 \rightarrow 0$ as $EW^2$
increases.

%%%
\PostScript{6}{0}{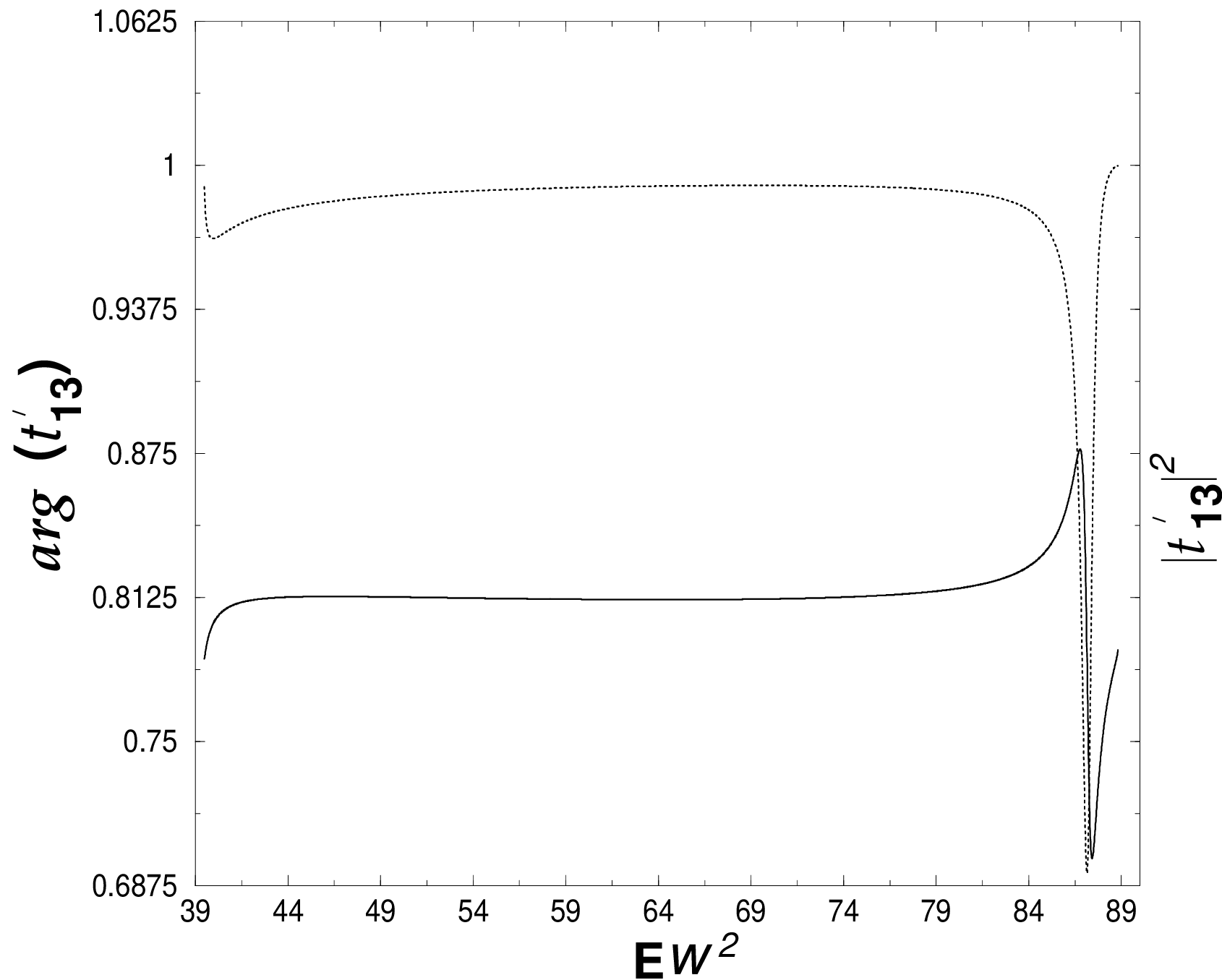}{0.5}{14}{
\hspace{-.5cm}
 The system under consideration is shown in Fig.7.  The solid curve gives
$arg(t_{13}^{'})$ in radians shifted by $\frac{\pi}{4}$ radians in the
y-direction and the dotted curve gives $|t_{13}^{'}|^2$.  Both the
functions are plotted
 versus $EW^2$ using $x_i=0$, $y_i=.45W$ and $\gamma=-10$
}{f2}
%%%
 \noindent Similarly if we rewrite Eq.(28) as
 \[ t_{24}^{'} = \frac{k_1(2\kappa_3+ g_3)+i\kappa_3 g_1}
{k_1(2\kappa_3+ g_3)+i\kappa_3(g_1+\beta g_2)}, \]
 where $\beta=\frac{k_1}{k_2}$;  then it is clear that the behavior of
$arg(t_{24}^{'})$ will be qualitatively the same.  It is indeed found in
Fig.8 that the behavior of $arg(t_{24}^{'})$ is similar to that of
$arg(t_{13}^{'})$.

$\th = \frac{1}{2i}ln[det[S]]$ is shown in Fig.12 as a function of energy,
for different values of $\gamma$. The minimum in $\th$ follows the
$E_{3b}$ and so the energy range where the slope of $\th$ versus E is
negative is determined by the $E_{3b}$. Note that when $E_{3b}$ goes out
of this energy range the $\th$ versus E has a positive slope everywhere.
So in Fig.12, the negative slope arises whenever a bound state $E_{3b}$ is
degenerate with the scattering states ($n$=1 and $n$=2), and
non-monotonously scatter and disperse the scattering states.
For weaker impurities in Q1D, the negative slope
occur at higher energies and also are steeper as demonstrated in Fig.12.
This is in contrast to what happens in 1D and demonstarted in Fig. 6,
that the energy where the negative slopes occur is always for
$E/V<1$.
 %%%
\PostScript{6}{0}{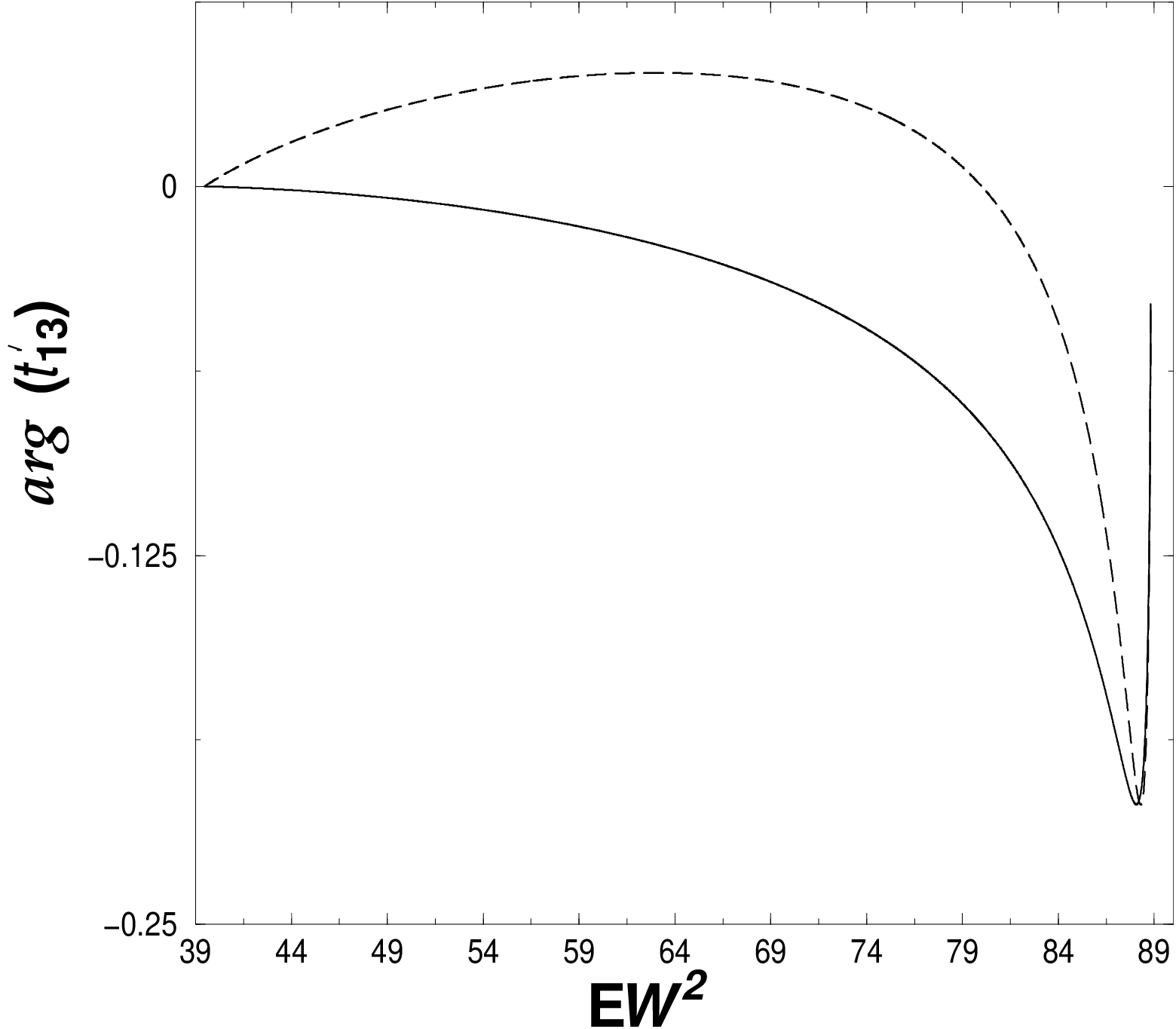}{0.5}{14}{
\hspace{-.5cm}
The system under consideration is shown in Fig.7.
The solid curve gives $arg(t_{13}^{'})$ in radians
shifted by $2\pi$ radians in the
negative y-direction versus $EW^2$ for
$\gamma=-47.1371$. The dashed curve is for $\gamma=-25.197$. We
use $x_i=0$ and $y_i=.21W$.
}{f2}
 %%%

%%
\PostScript{6}{0}{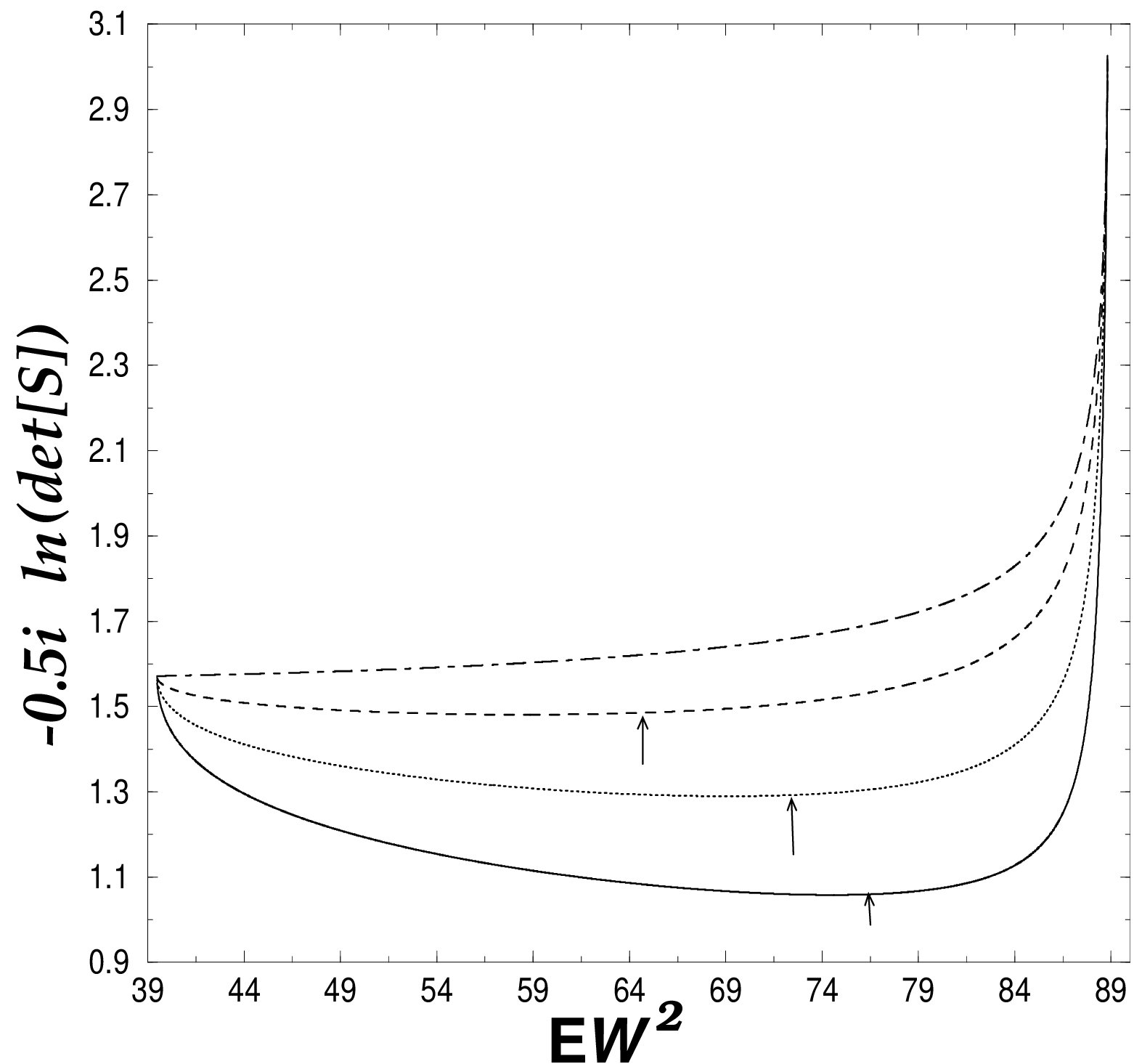}{0.5}{14}{
 The system under consideration is shown in Fig.7.  The plot is of $\th$
in radians versus $EW^2 $ for different $\gamma $. The dot-dashed curve is
shifted by $\pi$ radians in y-direction for $\gamma = -47.1371$,
corresponding $E_{3b}$ is at $EW^2=35$ which is less than the propagating
threshold $EW^2 \si 39$ of the second transverse mode.  The dashed curve
is for $\gamma = -25.197 $, corresponding $E_{3b}$ is at $EW^2=80$.  The
dotted curve is for $\gamma = -15$, corresponding $E_{3b}$ is at
$EW^2=86.606$.  The solid curve is for $\gamma = -10 $, corresponding
$E_{3b}$ is at $EW^2=87.982$. We use $ y_{i}=.21W$ and $x_i=0$. The arrows
accentuate the positions of the minima that is shifting towards higher
energies for weaker impurities.
 }{f2}
%%%

We have used two evanescent modes in our calculations because one can
include as many evanescent modes without changing the nature of the
negative slopes as long as the positions of the bound states $E_{3b}$ and
$E_{4b}$ remain the same. One can check this that with four evanescent
modes and $\gamma =-6.46584$, the negative slopes are the same as in Fig.
8, which means that the third and the fourth evanescent modes just
renormalizes $\gamma$ from -6.46584 to -10. The exact renormalization
takes place according to a formula $\gamma^h=\gamma^n/d$, where $\gamma^h$
is the $\gamma$ value used here and $\gamma^n$ is the renormalized value
of $\gamma$ when we use $m$ evanescent modes instead of two. $d=
(1+\Gamma_{55}^n/(2 \kappa_5) + \Gamma_{66}^n/(2 \kappa_6)....
\Gamma_{mm}^n/(2 \kappa_m)$, where $\Gamma_{mm}^n= =
\gamma^n Sin \frac{m \pi}{W}(y_{i}+\frac{W} {2})^2$.
Solving this one can find the renormalized value of $\gamma$ i.e.,
$\gamma^n$ that keep the minimum of any of the curves for the scattering
phase shifts considered here unchanged.  It is worthwhile mentioning that
at the band edges (i.e., $E \approx 39$ and 89, in the figures considered
in this section), the value of any curve is independent of the number of
evanescent modes, as all the modes get decoupled there.  In other words,
number of evanescent modes considered does not change the nature of the
negative slopes. Only the positions of the bound states are important.

In order to generalize to arbitrary number of propagating channels we
change our notations slightly.  For electrons incident from the
left/right, we call the scattered channels towards the left/right as
reflection channels (the rest being transmission channels) and change our
notation to
 \bea
{\tilde r_{11}} = t_{11}^{'}\hspace{.2cm}, \\
{\tilde  r_{22}} = t_{22}^{'}\hspace{.2cm}, \\
{\tilde r_{12}} = t_{12}^{'} .
\eea
 Thus all possible reflection channels are distinct. All the intra-subband
transmission channels are also distinct and they are denoted as ${\tilde
t_{11}}$ and ${\tilde t_{22}}$ where
 \bea
{\tilde t_{11}} = t_{13}^{'} \\
\mbox{and} \hspace{.5cm} {\tilde t_{22}} = t_{24}^{'} .
\eea
 All other scattering matrix elements are equal to one of these 5
elements.  In this notation the only difference is that the lowest channel
($n$=1)  on the transmission side is marked 1 instead of 3. So when we say
${\tilde t_{11}}$ we mean transmission amplitude from the $n$=1 channel on
the left to the $n$=1 channel on the right.  We find from Eqs. (29), (30),
(31) and (32)
 \be
\frac{d}{dE}\hspace{.1cm} arg({\tilde r_{mn}}) = \frac{d}{dE}
\hspace{.1cm}(\frac{1}{2i}
ln[det[S]])
\ee
 We find the above relation to be true for any number of propagating
modes.  So $m$ and $n$ can take any integer value less than or equal to
$p$, where $p$ is the total number of propagating modes. For two
propagating modes $p$=2, for three propagating modes $p$=3 and so on.  So
Eq.(41) is analogous to the 1D case given in Eq.(4).  That is when the
dimension of the matrix $S$ becomes very large, then it is sufficient to
consider the argument of a single matrix element in order to calculate the
complicated quantity on the RHS of Eq. 41.  In the energy regime where
there are two propagating channels, the negative slopes in $\th$ versus
incident energy curves are determined by $E_{3b}$, and when there are 3
propagating channels then the negative slopes are determined by $E_{4b}$
and so on.

The scattering phase shifts of transmission channels i.e. $arg({\tilde
t_{mn}})$, where again $m$ and $n$ can take all possible integer values
less than or equal to $p$, show sharp or gentle phase drops when the
scattering states are degenerate with a bound state, depending on the
value of the imaginary part in the numerator of ${\tilde t}_{mn}$. In the
single channel regime the imaginary part in the numerator is zero and
phase drops take the limiting value when the phase drops are absolutely
discontinuous by $\pi$. Just as the discontinuous phase drop in single
channel case do not affect $\th$ in any way, the phase drops of the
$arg({\tilde t}_{mn})$ also do not affect $\th$ in any way and $\th$
behaves similarly as $arg({\tilde r_{mn}})$.

\section{density of states and Friedel sum rule in quasi-one-dimension}

The local DOS is given by the following expression \cite{bag90}
\be
\rho_R = \int_{R} dx \int^{\frac{W}{2}}_{\frac{-W}{2}} dy
\sum_{m,k_m} \delta(E-E_{m,k_m}) \mid \psi_{m,k_m}(x,y)\mid ^2
\ee
 Here $E$ is the incident energy and R is the integration region where
modes are mixed. $m$ and $k_m$ are the two quantum numbers that define an
incident electron wavefunction, $A_me^{ik_mx}Sin
\frac{m\pi}{W}(y+\frac{W}{2})$ whose energy is $E_{m,k_m}$, where we have
taken that the electron is incident from the left i.e., $x<0$.
$\psi_{m,k_m}(x,y)$ is the wavefunction in the region of mode mixing and
$\psi_{m,k_m}(x,y)= \sum_n c_n^{(m)}
(x,k_n)Sin \frac{n\pi}{W}(y+\frac{W}{2})$.
Here $c_n^{(m)} (x,k_n)=C_ne^{ik_nx}$ for $n=1$ and $n=2$ and
$c_n^{(m)}(x,k_n)=C_ne^{-\kappa_nx}$ for $n > 2$; $x$ being greater than or
equal to 0.  The coefficients $C_n$ can be determined by using the mode
matching technique.  The mode matching has been done in details
by Bagwell \cite{bag90}. Here the delta function potential is taken to be
extending from $-\eps$ to $+\eps$ which has to be set to be tending to 0
in the end.  $\rho_{0R}$ can be determined by replacing
$\psi_{n,k_n}(x,y)$ by the plane wave states in absence of the scatterer
and doing the integration again.

Thus we find that for any non-zero incident energy
$$(\rho -\rho_0)_R =
\frac{2}{hv_1}[|t_{13}|^2+|t_{14}|^2+|t_{15}|^2+\cdots  ]$$
\be
+\frac{2}{hv_2}[|t_{23}|^2+|t_{24}|^2+|t_{25}|^2+\cdots  ].
\ee
 Here $v_1=\frac{\hbar k_1}{m}, v_2=\frac{\hbar k_2}{m}$ and
$t_{mn}=\frac{C_n}{A_m}$.  $t_{mn}$ can be obtained by solving the matrix
Eqs. given in Ref. \cite{bag90} (see Eq. 23 therein and we have used the
same notation i.e., $t_{mn}$ here is the same as $t_{mn}$ in Eq. (23)
of Ref. \cite{bag90}).
As can be seen in Fig.12 that
$\frac{d\th}{dE}$ is negative over a very large energy range
while $(\rho - \rho_0)_R$ as given by
Eq.43 is positive.

\PostScript{6}{0}{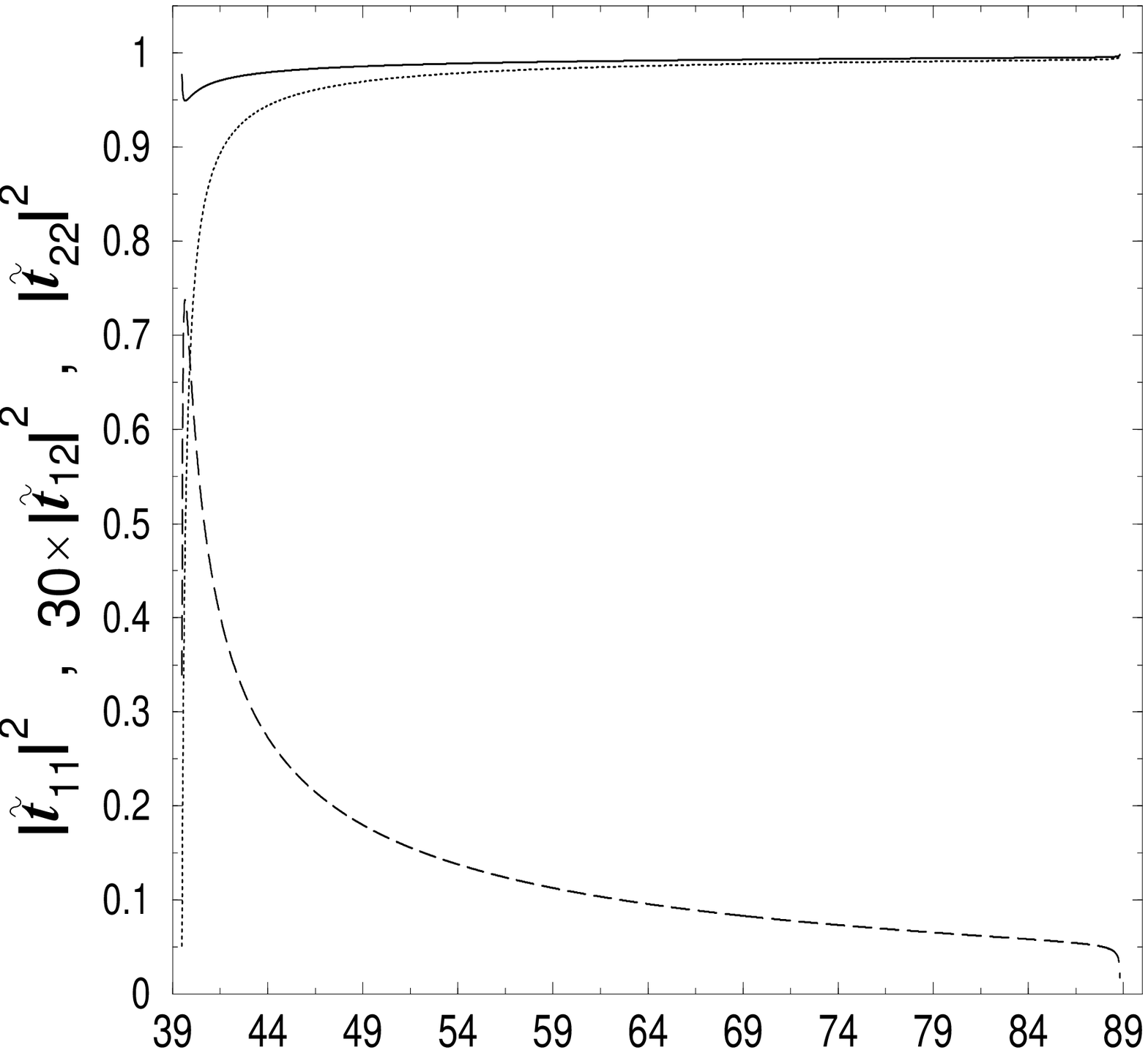}{0.5}{14}{
 The system under consideration is shown in Fig.7
with $\gamma = 1 $. The Fig. shows some important scattering
probabilities. The solid curve gives $|{\tilde t_{11}}|^2$
and it shows that
for $EW^2>50$, a particle incident in the first propagating
mode does not feel the scatterer at all, and is almost entirely
transmitted intra-channel, $|{\tilde t_{11}}|^2$ being close to unity.
The dotted curve gives $|{\tilde t_{22}}|^2$ and once again for
$EW^2>50$, it is close to unity signifying that a particle incident
in the second propagating channel is almost entirely transmitted
intra-channel. So $EW^2>50$ is the WKB regime where the potential
scatters the incident electron very weakly.
The dashed curve gives 30 times $|{\tilde t_{12}}|^2$
and shows strong energy dependence not only for $EW^2<50$ but also
around the highest energy ($EW^2 \approx 89$)
or in the extreme WKB limit, its
absolute value being extremely small there signifying extremely low
inter-channel transmission i.e., the incoming particle does not
feel the scatterer.
 }{f2}
%%%

%%
\PostScript{6}{0}{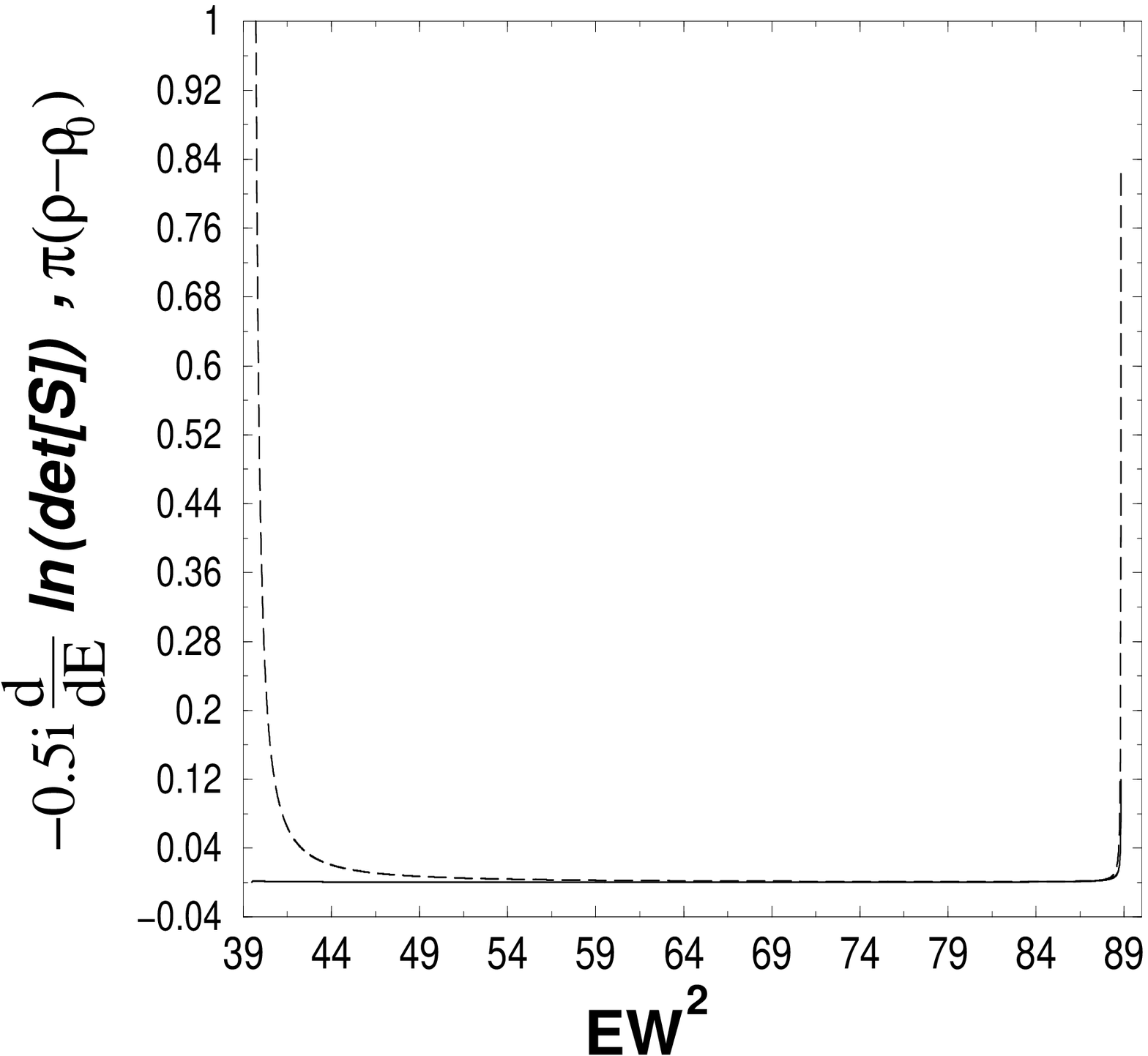}{0.5}{14}{
 The system under consideration is shown in Fig.7 with $\gamma=1$.
The solid curve gives $\pi (\rho(E) - \rho_0(E))$ and the dashed
curve gives -0.5$i {d \over dE} ln(det[S])$. The two curves
deviate from each other, where ever the curves in Fig. 13 are
strongly energy dependent. Otherwise they agree.
 }{f2}
%%%

Note that if we calculate the global DOS by taking the integration
region to be from $-\infty$ to $\infty$ instead of just the region
R where the modes are mixed then Eq. 43 remain the same. One will
get some extra integrals that are indefinite integrals but using
the current conservation condition it can be analytically proved
(see appendix II)
that they cancel each other. This means the contribution to $\rho(E)$
and that to $\rho_0(E)$ coming from outside the region R cancel each
other in the absence of a term like $\rho_q/l$ as in section II.
Thus in this case (the proof is given in appendix II)
 $$\rho(E)-\rho_0(E)=(\rho-\rho_0)_R$$
 and both of them deviate identically from ${1 \over \pi}{d \theta \over
dE}$ due to strong dispersion, at any arbitrary energy.

Since the negative slopes are due to the bound states supported
by the negative delta function potential, one may ask what happens for
a positive delta function potential that does not support any bound
state. This situation is discussed below and it
also elaborates the uniqueness of the Q1D,
with respect to the violation of Friedel sum rule and shows that
large violation can occur in the extreme WKB limit. Fig 13 shows
the energy dependence of some important scattering probabilities.
As in 1D the scattering probabilities are strongly energy dependent
for $EW^2<50$. For $EW^2>50$ all
the curves vary slowly with energy (non-dispersive scattering).
However,
one scattering probability $|{\tilde t_{12}}|^2$ is also very
strongly energy dependent at the highest energy of $EW^2 \approx 89$.
($k_m^2/V>>1$). In this regime scattering is almost entirely intra-channel
as can be seen that $|{\tilde t_{11}}|^2$ and $|{\tilde t_{22}}|^2$ are
both almost unity. All other scattering probabilities like reflection
probabilities and inter-channel scattering probabilities like
$|{\tilde t_{12}}|^2$ are extremely small. There are several other
scattering matrix elements that are identical to ${\tilde t_{12}}$.
Although being small,
they can have strong energy dependence, or their energy derivative
can be very large. As can be seen in Fig. 13 that the dashed curve
bends down with a steep slope at $EW^2 \approx 89$.
It is found in Fig. 14 that there is a strong violation
of Friedel sum rule in the non-WKB regime as in 1D and also in the
extreme WKB regime ($EW^2 \approx 89$ where $k^2/V>>1$ and scattering
is almost completely intra-channel)
quite unlike that in 1D. It can be seen in Fig. 13
that in the mid-energy range, the scattering probabilities are
not very energy dependent and the Friedel sum rule holds in this
mid-energy range as can be seen in Fig. 14. But at regimes
where the scattering leads to strong dispersion, FSR is violated.
As the strength of the
positive delta function potential is increased, this regime where
the scattering probabilities are not strongly energy dependent becomes
narrower and Friedel sum rule is violated at all energies.

%%%%%%%%%%%%%%%%%%%%%%%%%%%%%%%
\section{ phase behavior at critical energies }
\label{s4}
%%%%%%%%%%%%%%%%%%%%%%%%%%%%%%%

Very interesting phase behaviors can be seen at energies where the
$S$-matrix changes dimension. For example for $E\le \frac{4\pi ^2}{W^2}$
there is only one propagating mode and the $S$-matrix is 2$\times$2. But
for $E > \frac{4\pi ^2}{W^2}$, there are two propagating modes and the
$S$-matrix is 4$\times$4. The matrix element $t_{11}^{'}$ exists on either
side of the energy $\frac{4\pi ^2}{W^2}$ and in Fig.15 we show the
behavior of $arg(t_{11}^{'})$ in the energy range that includes $EW^2 =
4\pi ^2 $.  Note that it exhibits a discontinuous phase drop by
$\frac{\pi}{2}$ at $EW^2 = 4\pi ^2 $. So far only discontinuous phase
drops of $\pi$ has been observed but never $\frac{\pi}{2}$.
 From the properties of a 2$\times$2 $S$-matrix it follows that if there
is a discontinuous phase change then it can only be of $\pi$ \cite
{lee99,tan99}. So had the $S$-matrix been 2$\times$2 on either side of
$EW^2 = 4\pi ^2$ the phase drop would have been $\pi$. But since the
$S$-matrix is 2$\times$2 only on one side, including $EW^2 = 4\pi ^2$,
i.e., $E\le \frac{4\pi ^2}{W^2}$, the phase drop is also one half of
$\pi$. $|t_{11}^{'}|^2$ also has a zero at $EW^2 = 4\pi ^2$ for all
possible choice of parameters \cite{bag90}, and this zero is associated
with a $\frac{\pi}{2}$ phase jump instead of a $\pi$ phase jump.

%%%
\PostScript{6}{0}{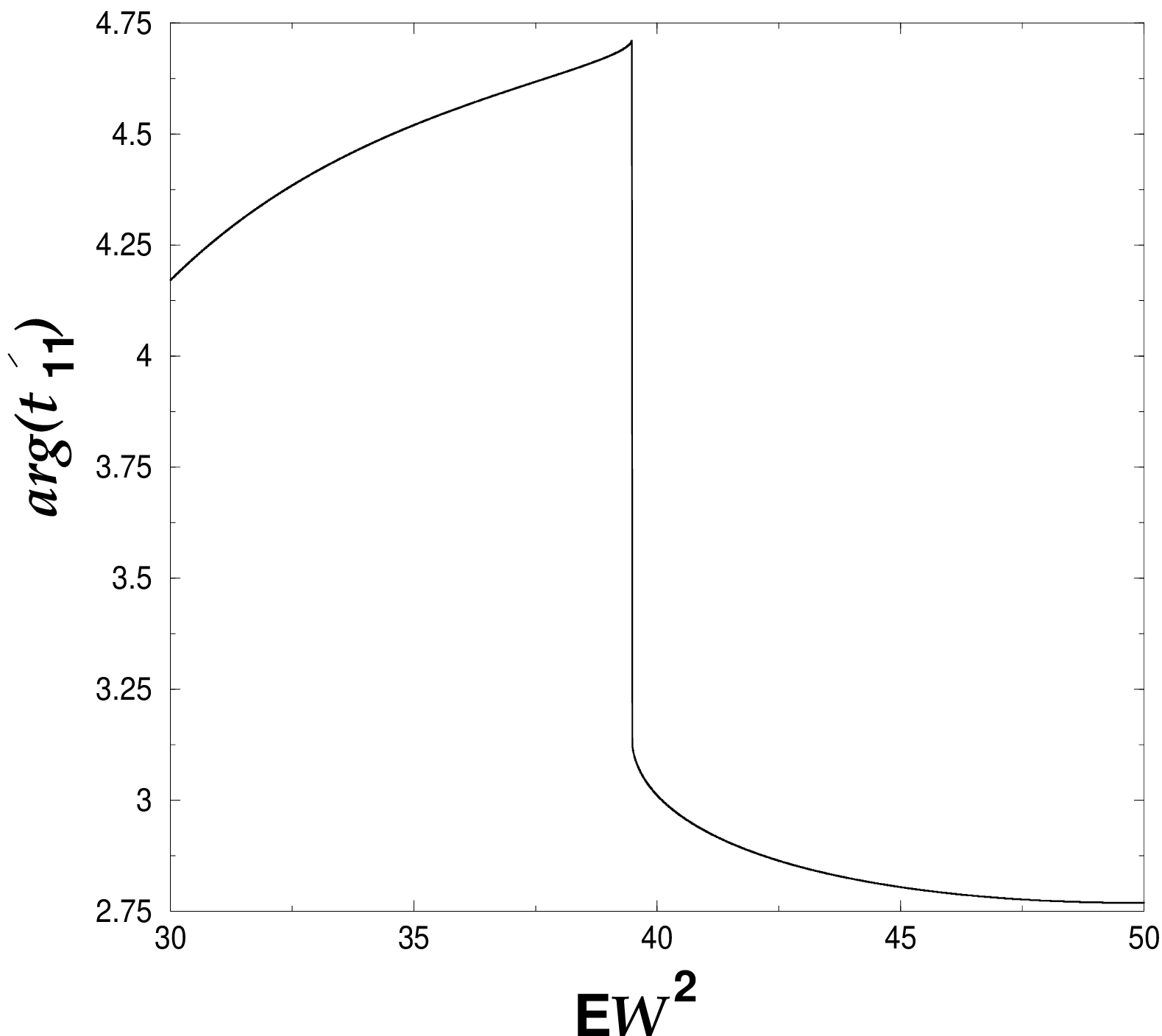}{0.5}{14}{
\hspace{-.5cm}
 The system under consideration is shown in the Fig.7.  The plot is of
$arg(t_{11}^{'})$ in radians versus E$W^2$.  This plot is for
$\gamma=-25.197$, $x_i=0$ and $y_i=.45W$.
 }{f2}
%%%

%%%
\PostScript{6}{0}{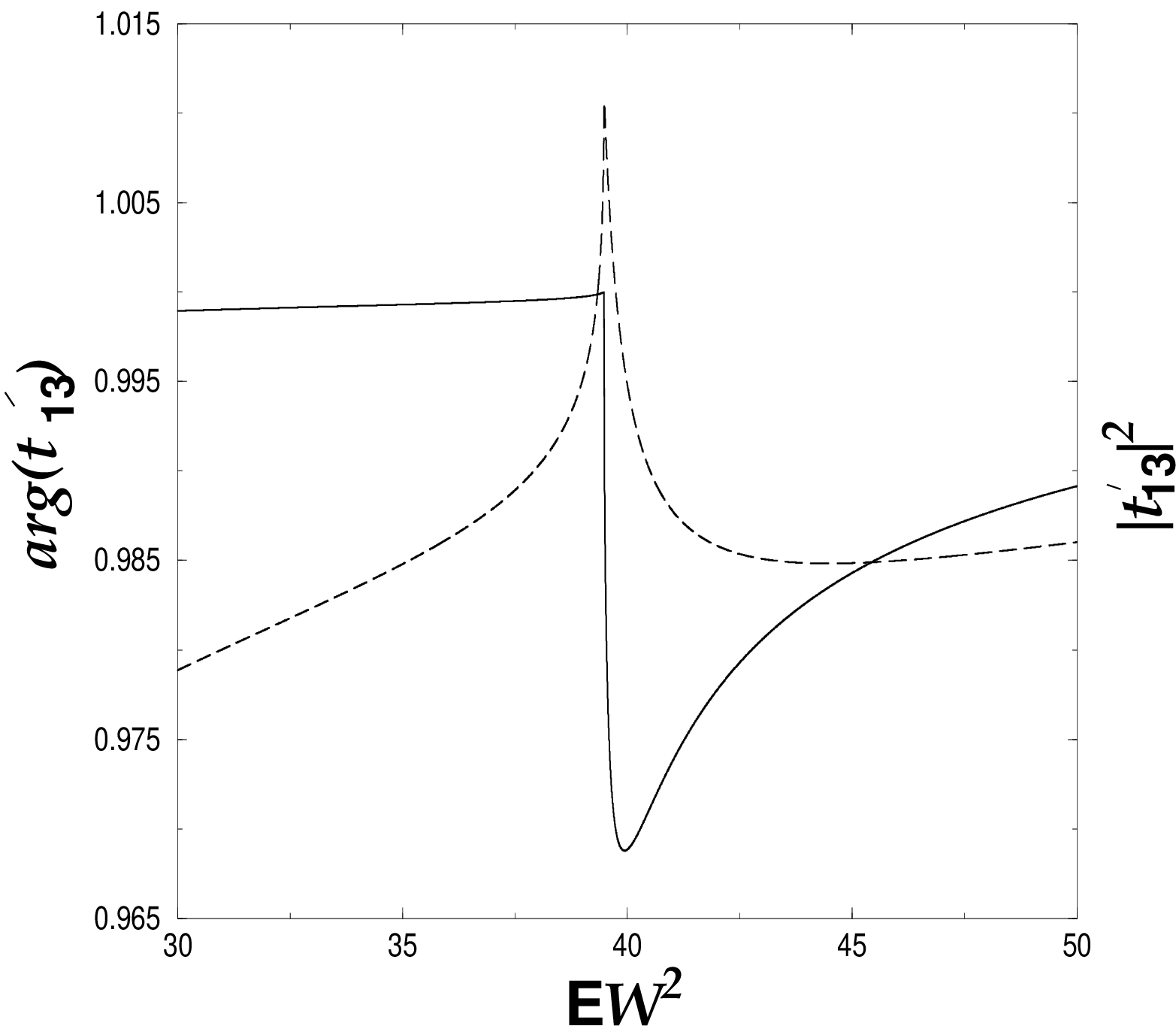}{0.5}{14}{
\hspace{-.5cm}
 The system under consideration is shown in the Fig.7.  The solid curve
gives $|t_{13}^{'}|^2$. The dashed curve is after subtracting 5.27183
radians from $arg(t_{13}^{'})$ in radians.  We use $\gamma=+25.197$,
$x_i=0$ and $y_i=.45W$.
 }{f2}
%%%

Next we take a repulsive $\delta$ function potential. It is known
\cite{bag90} that at critical energies like $EW^2=4\pi^2$,
$|t_{13}^{'}|^2$ shows discontinuities.  Here $|t_{13}^{'}|^2$ does not
have a zero but exhibits a discontinuous jump.  At these points
$arg(t_{13}^{'})$ also shows non-analytic behavior as demonstrated in
Fig.16. In this case $\frac{d}{dE} arg(t_{13}^{'})$ is discontinuous.

%%%%%%%%%%%%%%%%%%%%%%%%%%%%%%%
\section{Conclusions}
\label{s5}
%%%%%%%%%%%%%%%%%%%%%%%%%%%%%%%

In a multichannel quantum wire with attractive impurities, negative slopes
in the scattering phase shift versus incident energy curves can occur at
all possible energies. For weaker defects it happens at higher energies
and the negative slopes are more pronounced.
Such negative slopes
mean super luminescence \cite{smi59,kumar} that can be observed
experimentally. Although such a super luminescent particle will not give
any information about the particle delay or information delay, they are of
interest because they demonstrate fundamental principles in quantum
mechanics.  Hence Eq. (41) derived in this paper may be of use to
experimentalists and theoreticians.  Calculations of variation of DOS due
to the presence of the scatterer show that FSR can be violated at any
arbitrary energy. That is violation of FSR is not restricted to 
non-WKB regime (the regime where anyway transport does not occur)
as in 1D, 2D or 3D.
Rather the violation is related to strong energy dependence of the
scattering matrix elements (dispersive behavior). 
While in 1D, 2D and 3D the strong
energy dependence of scattering occur only in the non-WKB regime,
in Q1D there is no systematics. A Q1D system in the extreme
WKB limit can also exhibit strongly energy dependent scattering
and there is no definable regime where the FSR will work and
hence it may not work even in regimes where transport occurs.
For attractive impurities, weaker the impurity, stronger the
violation of FSR while for repulsive impurities, stronger the
impurity, stronger the violation of FSR.
We also show that the discontinuous phase drops in the
single channel case have a counterpart in the multichannel case wherein
the drops can be continuous and we propose a line shape formula for them
in Eq. 34. When there is a third channel of escape
for the electron, apart from the channel along which
it is incident and the channel where its scattering
phase shift is measured, the phase drop becomes
continuous. However, these phase drops do not
affect ${1 \over 2i} ln Det [S]$ and hence Friedel sum rule.
Finally, we discuss some novel scattering phase shifts at
energies where the $S$ matrix changes dimension.

\section{ACKNOWLEDGMENTS}

We thank Dr. A. M. Jayannavar for useful discussions.

\section{APPENDIX I}

Considering the symmetric scattering potential in Fig.4, the scattering
matrix $ S $ of the structure can be found by cascading the scattering
matrices of different parts, i.e.,\\
 \[ S =  \left(\begin{array}{cc}
\displaystyle r & \displaystyle t \\
\displaystyle t & \displaystyle r
\end{array}\right) =S_{1} \otimes  S_{2} \otimes  S_{3} ,\]
\[ \mbox{where} \hspace{.4cm} S_{1} = S_{3} = \left(\begin{array}{cc}
\displaystyle r' & \displaystyle t' \\
\displaystyle t' & \displaystyle r'
\end{array}\right) \]
\hfill\\
\[ \mbox{and} \hspace{.4cm} S_{2} = \left(\begin{array}{cc}
0 & \displaystyle \tau \\
\displaystyle \tau & 0
\end{array}\right)\hspace{.1cm}.\]
 Here $\tau = e^{i \phi}$, $\phi = kl$ and $k = \sqrt{\frac{2m}{\hbar^{2}}
E }$.  $S_{2}$ is the scattering matrix for the free region II of length
$l$ between the two scatterers. $ r' $ \& $ t' $ are the reflection \&
transmission amplitudes due to one of the two potentials when isolated.

After cascading these three matrices the resultant scattering matrix of
our system becomes
 \[  S = \left(\begin{array}{cc}
\displaystyle r'+\frac{t'^{2}{\tau}^{2}r'}{1-r'^{2}{\tau}^{2}} &
\displaystyle \frac{t'^{2}{\tau}}{1-r'^{2}{\tau}^{2}}\\
\displaystyle \frac{t'^{2}{\tau}}{1-r'^{2}{\tau}^{2}} &
\displaystyle r'+\frac{t'^{2}{\tau}^{2}r'}{1-r'^{2}{\tau}^{2}}
\end{array}\right) \hspace{.1cm}. \]
And so,
\[ det [S] = \left(\displaystyle r'+\frac{t'^{2}{\tau}^{2}r'}{1-r'^{2}{\tau}
^{2}}\right)^{2} - \left(\displaystyle \frac{t'^{2}{\tau}}{1-r'^{2}{\tau}^{2}}
\right)^{2} \hspace{.5cm} \mbox{(i)}\]
\[   \hspace{1cm}= \frac{1}{(1-r'^{2}{\tau}^{2})^2} \hspace{.1cm} (M+N)
\hspace{.1cm},\hspace{2.5cm} \mbox{(ii)}\]
\[ \mbox{where}\hspace{1cm}M = r'^{2}(1-r'^{2}{\tau}^{2})^2 \hspace{2.0cm}
\mbox{(iii)} \]
\[ \mbox{and} \hspace{.5cm} N = (1-r'^{2}{\tau}^{2})\hspace{.1cm} (2t'^{2}r'^{2}
{\tau}^{2}-t'^{4}{\tau}
^{2}) . \hspace{1cm} \mbox{(iv)}  \]
From (i),
\[ \frac{\pa det [S] }{\pa \phi} = 2\left( A \frac{\pa A}{\pa \phi} -B
\frac{\pa B}{\pa \phi} \right) , \hspace{2.0cm}\mbox{(v)}\]
\[ \mbox{where}\hspace{1cm} A = r'+\frac{t'^{2}
{\tau}^{2}r'}{1-r'^{2}{\tau}^{2}} \hspace{2cm}\mbox{(vi)}\]
\[ \mbox{and} \hspace{.5cm} B =
\frac{t'^{2}{\tau}}{1-r'^{2}{\tau}^{2}}\hspace{.5cm}
.\hspace{2cm}\mbox{(vii)}\]
Using (v), (vi) and (vii) we find
$$  \hspace{-5cm}\frac{\pa det [S] }{\pa \phi} = $$
\[  \hspace{1.0cm} 2\left[A
\left(2r'B+\frac{2r'^{3}{\tau}^{2}
B}{1-r'^{2}{\tau}^{2}}\right) -
B\left(\frac{t'^{2}+r'^{2}t'^{2}{\tau}^{2}}
{(1-r'^{2}{\tau}^{2})^2}\right)\right] \frac{\pa \tau }{\pa \phi} \]
\[ \hspace{1cm} + 2\left[A
\left(1+B{\tau}+\frac{2r'^{2}
{\tau}^{3}B}{1-r'^{2}{\tau}^{2}}\right)-B
\left(\frac{2r'{\tau}^{2}B}{1-r'^{2}{\tau}^{2}}\right)\right] \frac{\pa r'}{\pa \phi} \]
\[ \hspace{1cm} + 2\left[2r'{\tau}AB
-B\left(\frac{2{\tau}t'}
{1-r'^{2}{\tau}^{2}}\right)\right] \frac{\pa t'}{\pa \phi} .\hspace{1.5cm} \mbox
{(viii)} \]

We will neglect the last two terms in comparison to the 1st term in
Eq.(viii) but we will retain the energy dependence of $ r' $ and $ t' $
and the 1st term in Eq.(viii) is actually varying with energy very
strongly. Hence our results correspond to real potentials and we do not
parameterize the $ S $ matrix in a special way. We apply this result to
the case of double delta function potential in Fig.5 and illustrate the
significance of the last two terms in comparison with the 1st one. We
stress that this calculation in this appendix holds even if the $\delta
$-function potential is replaced by the square-well or any arbitrary
potential.  Thus in the regime where $ \frac{\pa r'}{\pa \phi} \to 0 $ and
$\frac{\pa t'}{\pa
 \phi} \to 0 $ and using $\frac{\pa \tau }{\pa \phi} = i\tau$,
 $$ \hspace{-7cm} \frac{\pa det [S] }{\pa \phi} $$
\[ = 2\left[A\left(2r'B+\frac{2r'^{3}{\tau}^{2}
B}{1-r'^{2}{\tau}^{2}}\right) -
B\left(\frac{t'^{2}+r'^{2}t'^{2}{\tau}^{2}}
{(1-r'^{2}{\tau}^{2})^2}\right)\right] i\tau  \hspace{.2cm}. \]
At this point we substitute the values of $A$ and $B$ from (vi) and (vii)
to get
\[ \frac{\pa det [S] }{\pa \phi} = 2i\hspace{.2cm}
\frac{1}{(1-r'^{2}{\tau}^{2})^3}\hspace{.2cm}N \hspace{.2cm}. \hspace{1.5cm}
\mbox{(ix)}\]
\[ \mbox{Now},\hspace{.2cm} \th = \frac{1}{2i} ln (det [S] )\hspace{4cm}
\mbox{(x)} \]
\[ \mbox{and so} \hspace{.4cm} \frac{\pa \th}{\pa E} = \frac{\pa \th}{\pa
\phi} \hspace{.2cm}
\frac{\pa \phi}{\pa E} . \]
From (x), (ii) and (ix)\\
\[ \frac{\pa \th}{\pa \phi} = \frac{1}{2i} \hspace{.2cm} \frac{1}{det [S] }
\hspace{.2cm} \frac{\pa det [S] }{\pa \phi} \]
\[  \hspace{1.5cm} = \frac{1}{1-r'^{2}{\tau}^{2}}\hspace{.2cm}\frac{1}
{\frac{M}{N}+1} \hspace{.2cm}. \]
Multiplying the numerator and denominator by $\frac{(1-r'^{2}{\tau}^{2})^{*}}
{1-\mid r^{'} \mid^{4}}$, we get
\[  \hspace{.2cm}  \frac{\pa \th}{\pa \phi} = \frac{1-\mid r' \mid ^{4}}
{\mid 1-r'^{2}{\tau}^{2} \mid
^{2}} \hspace{.2cm} \frac{(1-r'^{2}{\tau}^{2})^{*}}{1-\mid r' \mid ^{4}}
\hspace{.2cm}\frac{1}{\frac{M}{N}+1} \]
\[  \hspace{.2cm}  = \frac{1-\mid r' \mid ^{4}}{\mid 1-r'^{2}{\tau}^{2} \mid
^{2}} \hspace{.2cm}Q \hspace{.2cm},\hspace{4cm}\mbox{(xi)}\]
\[ \mbox{where,}  \hspace{.5cm} Q=\frac{(1-r'^{2}{\tau}^{2})^{*}}{1-\mid r'
\mid ^{4}}
\hspace{.2cm}\frac{1}{\frac{M}{N}+1} \hspace{2cm} \mbox{(xii)}\]
\[ \hspace{.2cm} =\left[\frac{(1-{r'^{*}}^{2}{{\tau}^{*}}^{2})-1+\mid
r'\mid^{4}\mid
\tau \mid^{4}}{1-\mid r' \mid^{4}}+1\right]\hspace{.2cm}
\frac{1}{\frac{M}{N}+1} \hspace{.2cm},\]
\[ \hspace{5cm} \mbox{as}\hspace{.4cm} |\tau |^4 = 1\hspace{.2cm}. \]
As \hspace{.5cm} $1-|r'|^2=|t'|^2$
\[ \hspace{.2cm} Q =\left[\frac{-{r'^{*}}^{2}{{\tau}^{*}}^{2}+\mid r'\mid^{4}
\mid \tau \mid^{4}}{(1+\mid r' \mid^{2})|t'|^2}+1\right]\hspace{.2cm}\frac{1}
{\frac{M}{N}+1}  \hspace{.2cm}.\]
Now substituting the values of $M$ and $N$ from (iii) and (iv)
\[ Q = \frac{\frac{-{r'^{*}}^2{{\tau}^{*}}^2(1-r'^{2}
{\tau}^{2})+\mid t' \mid ^{2}(1+\mid r' \mid ^{2})}{\mid t' \mid ^{2}
(1+\mid r' \mid ^{2})}}{
\frac{r'^{2}(1-r'^{2}{\tau}^{2})-t'^{2}{\tau}^{2}(t'^{2}-2r'^{2})}{-t'^{2}
{\tau}^{2}(t'^{2}-2r'^{2})}} \hspace{.2cm}.\]
\hfill\\
Using, \hspace{.4cm} $  r' = \mid r' \mid e^{i\th _{r}} $ \hspace{.4cm} and
\hspace{.4cm} $ t'= \mid t' \mid e^{i\th _{t}} $,
$$Q=[|r'|^2|t'|^2|\tau |^4e^{2i\th _{t}}e^{2i(\th _{t}-\th _{r})}
-|r'|^4|t'|^2|\tau
|^4\tau ^2e^{4i\th _{t}}$$
$$-|t'|^4\tau^2e^{4i\th _{t}}
-|t'|^4|r'|^2\tau^2e^{4i
\th _{t}}$$
$$-2|r'|^4|\tau |^4e^{2i\th _{t}}
+2|r'|^6|\tau |^4\tau^2e^{2i(\th _{t}
+\th _{r})}$$
$$+2|r'|^2|t'|^2\tau^2e^{2i(\th _{t}+\th _{r})}+2|r'|^4||t'|^2\tau^2
e^{2i(\th _{t}+\th _{r})}]/D ,$$
\[\mbox{where},\hspace{.2cm} D = (1+\mid r' \mid^{2})(-\mid t' \mid ^{4}{\tau}
^{2}e^{4i\th _{t}}-\mid r' \mid ^{4}{\tau}^{2}e^{4i\th _{r}}\]
\[ \hspace{.2cm} +2\mid t' \mid ^{2}\mid r' \mid ^{2}\tau ^2 e^{2i
(\th _{r}+\th _{t})}+\mid r' \mid ^{2}e^{2i\th _{r}}) . \hspace{1cm}
\mbox{(xiii)}\]
\hfill\\
Now it follows from unitarity that
$e^{2i(\th _t - \th _r)}  =  e^{i \pi} = -1$ and
$|\tau |^4 = 1$ and so
\hfill\\
\[ Q = [-|r'|^2 |t'|^2  e^{2i \th _t } - |r'|^4 |t'|^2 \tau ^2 e^{4i \th _t }
-|t'|^4 \tau ^2 e^{4i \th _t } \]
\[ - |t'|^4 |r'|^2 \tau ^2 e^{4i \th _t } - 2
|r'|^4 e^{2i \th _t } + 2 |r'|^6 \tau ^2  e^{2i ( \th _t +  \th _r)} \]
\[ + 2 |r'|^2 |t'|^2 \tau ^2  e^{2i ( \th _t +  \th _r)} +  2 |r'|^4 |t'|^2
\tau ^2  e^{2i ( \th _t +  \th _r)}]/D \]
\[\hspace{1cm} = [- e^{2i \th _t } |r'|^2 \{|t'|^2 + 2 |r'|^2 \} - (|t'|^2 +
|r'|^2 \{|r'|^2 + |t'|^2\}) \]
\[ (e^{4i \th _t } |t'|^2  \tau ^2 -  e^{2i ( \th _t +
\th _r)} 2 |r'|^2 \tau ^2 )]/D \hspace{.2cm}.\]
Now inside the $\{  \} $ brackets if we use the fact that $ \mid r' \mid ^{2}+
\mid t' \mid ^{2}=1$, then\\
\[ Q = [- e^{2i \th _t } |r'|^2 (1+|r'|^2) - e^{4i \th _t }|t'|^2 \tau ^2
+ 2 e^{2i (\th _t + \th _r)}  |r'|^2 \tau ^2]/D .\]
Multiplying numerator and denominator above by $e^{-2i(\th _{r}+\th _{t})}$ and
putting $ e^{2i (\th _t - \th _r)} = e^{i \pi} = -1$, we get
\[ Q = \frac{-e^{-2i \th _{r}} |r'|^2 (1+|r'|^2 )+\tau ^2 (|t'|^2 +|r'|^2)+
|r'|^2
\tau ^2}{D'} ,\]
\[ \mbox{where,}\hspace{.1cm} D'= De^{-2i(\th _{r}+\th _{t})} .\hspace{2cm}
\mbox{(xiv)} \]
Again using $|r'|^2 + |t'|^2 =1$,
\[ Q = \frac{(1+|r'|^2 ) (\tau ^2 - |r'|^2 e^{-2i\th _{r}})}{D'} . \hspace{1cm}
\mbox{(xv)} \]
Now from (xiv) and (xiii)
\[ D' = (1+|r'|^2)[|r'|^2  e^{-2i  \th _t} - |r'|^4 \tau ^2  e^{-2i(\th _t -
\th _r)}  \]
\[-|t'|^4 \tau ^2 e^{2i(\th _t - \th _r)} + 2|t'|^2 |r'|^2 \tau ^2] . \]
As $\th _t - \th _r = \frac{\pi}{2}$,
\[ \hspace{2cm} D' = (1+|r'|^2)[|r'|^2  e^{-2i  \th _t} + \tau ^2 (|r'|^2 +
 |t'|^2 )^2 ] \]
where of course $\mid r' \mid^{2}+\mid t' \mid ^{2}=1$.
Substituting $D'$ in Eq.(xv),
\[ Q = \frac{{\tau}^{2} - \mid r' \mid ^{2}e^{-2i\th _{r}}}{{\tau}^{2}+\mid
r' \mid ^{2}e^{-2i\th _{t}}} .\]
Multiplying numerator and denominator of $Q$ by
$e^{2i\th_{t}}$ and using $e^{2i
(\th _t - \th _r)} = e^{i \pi} = -1$,
we get from (xi)
\[  \frac{\pa \th}{\pa \phi}  =\frac{1-\mid r' \mid ^{4}}{\mid 1-r'^{2}{\tau}
^{2} \mid
^{2}} \]

\section{APPENDIX II}

The global density of states is given by
\bea
\rho(E)&=&\int_{-\infty}^\infty dx \int_{\frac{-W}{2}}^{\frac{W}{2}}dy\nn\\
&& \sum_{m,k_m}\left|\psi_{m,k_m}(x,y)\right|^2 \delta (E-E_{m,k_m})
\hspace{1.6cm}\mbox{(i)}\nn
\eea
where \hspace{0.2cm} $\psi_{m,k_m}(x,y) = \sum_n c_n^{(m)}(x)\chi_n(y)$
and $E_{m,k_m}$ is the energy of an electron in the leads.\\
\hfill\\
$E_{m,k_m}=\displaystyle \frac{m^2\pi^2\hbar ^2}{2m_eW^2}+ \frac{\hbar ^2k_m^2}{2m_e},\hspace{0.2cm}\mbox{where}\hspace{0.2cm}m=\pm 1, \pm 2,$
as there are two propagating modes in the leads.\\
\hfill\\
As $\chi_n(y)$'s form an orthonormal set,
$$
\rho(E)=\sum_{m,k_m}\delta (E-E_{m,k_m})\int_{-\infty}^\infty
dx \sum_n \left|c_n^{(m)}(x)\right|^2$$
\bea
\rho(E)&=&\frac{2}{h v_1}\int_{-\infty}^\infty dx
\sum_n \left|c_n^{(1)}(x)\right|^2\nn\\
&+& \frac{2}{h v_2}\int_{-\infty}^\infty dx \sum_n \left|c_n^{(2)}(x)\right|^2
\hspace{2.5cm}\mbox{(ii)}\nn
\eea
Here, $v_1 = \displaystyle \frac{\hbar k_1}{m_e}$ and $v_2 =
\displaystyle \frac{\hbar k_2}{m_e}$.\\
\hfill\\
Now,
\bea
\int_{-\infty}^\infty dx \sum_n \left|c_n^{(1)}(x)
\right|^2 &=& \int_{-\infty}^\infty \left|c_1^{(1)}(x)\right|^2 dx \nn\\
&+& \int_{-\infty}^\infty \left|c_2^{(1)}
(x)\right|^2 dx\nn\\
&+& \int_{-\infty}^\infty \left|c_3^{(1)}(x)\right|^2 dx +
\cdots  \nn\\
&=& T1 \mbox{(say)},\nn
\eea
where electron is incident in the fundamental mode,
\bea
c_1^{(1)}(x) &=& e^{ik_1x} +
{\tilde r_{11}}  e^{-ik_1x}
\hspace{.2cm}\mbox{for}\hspace{.2cm} x < 0 \nn\\
&=& {\tilde t_{11}} \hspace{.2cm}e^{ik_1x}
\hspace{.2cm}\mbox{for}\hspace{.2cm}
x > 0 \nn\\
c_2^{(1)}(x) &=& {\tilde r_{12}}
\hspace{.2cm}e^{-ik_2x}
\hspace{.2cm}\mbox{for}\hspace{.2cm} x < 0 \nn\\
&=& {\tilde t_{12}} \hspace{.2cm}e^{ik_2x}
\hspace{.2cm}\mbox{for}\hspace{.2cm}
x > 0 \nn
\eea
and for $n > 2$,
\bea
c_n^{(1)}(x) &=& t_{1n} \hspace{.2cm}e^{\kappa_n x}\hspace{.2cm}
\mbox{for}\hspace{.2cm}x < 0 \nn\\
&=&  t_{1n} \hspace{.2cm}e^{-\kappa_n x}\hspace{.2cm}\mbox{for}
\hspace{.2cm}x >
0 \nn
\eea
\bea
T1 &=& \int_{-\infty}^0 dx
\left[1+|{\tilde r_{11}}|^2+2|{\tilde r_{11}}| \hspace{.2cm}Cos(2k_1x+\eta_1)
\right] \nn\\
 &+& \int_0^\infty dx|{\tilde t_{11}}|^2
+ \int_{-\infty}^0 dx|{\tilde r_{12}}|^2 +
\int_0^\infty dx|{\tilde t_{12}}|^2\nn\\
&+&|t_{13}|^2 + |t_{14}|^2 + \cdots \cdots \nn
\eea
Here,\hspace{.3cm}$\eta_1$ is defined as $\displaystyle{\tilde r_{11}}=|{\tilde r_{11}}|e^{-i\eta_1}$\\
\hfill\\
Similarly, for electron incident in the first excited mode,\\
\bea
T2 &=& \int_{-\infty}^\infty dx \sum_n \left|c_n^{(2)}(x)\right|^2\nn\\
&=& \int_{-\infty}^0 dx \left[1+|{\tilde r_{22 }}|^2+2|{\tilde r_{22}}| \hspace{.2cm}Cos(2k_2x+\eta_2)\right]\nn\\
&+& \int_0^\infty dx|{\tilde t_{22}}|^2 + \int_{-\infty}^0 dx|{\tilde r_{21}}|
^2 + \int_0^\infty dx|{\tilde t_{21}}|^2\nn\\
&+&|t_{23}|^2 + |t_{24}|^2 + \cdots \cdots  ,\nn
\eea
Here,\hspace{.3cm}$\eta_2$ is defined as $\displaystyle{\tilde r_{22 }}= |{\tilde r_{22 }}| e^{-i\eta_2}$\\
\bea
\mbox{Therefore,}\nn\\
\rho (E)&=& 2\left[\frac{1+|{\tilde r_{11}}|^2}{h v_1}\int_{-\infty}^0 dx \hspace{.2cm}+\hspace{.2cm} \frac{1+|{\tilde r_{22}}|^2}{h v_2}\int_{-\infty}^0 dx \right.\nn\\
&+& \left. \frac{|{\tilde r_{12}}|^2}{h v_1}\int_{-\infty}^0 dx +
\hspace{.02cm}\frac{|{\tilde r_{21}}|^2}{h v_2}\int_{-\infty}^0 dx \right.\nn\\
 &+& \left. \frac{2|{\tilde r_{11}}|}{h v_1}\int_{-\infty}^0 dx \hspace{.2cm}Cos(2k_1x+\eta_1)\right. \nn \\
& + & \left. \hspace{.02cm}\frac{2|{\tilde r_{22}}|}{h v_2}\int_{-\infty}^0 dx \hspace{.2cm}Cos(2k_2x+\eta_2)\right. \nn \\
&+& \left. \frac{|{\tilde t_{11}}|^2+|{\tilde t_{12}}|^2}{h v_1}\int_0^{\infty}dx + \frac{|{\tilde t_{22}}|^2+|{\tilde t_{21}}|^2}{h v_2}\int_0^{\infty}dx \right. \nn \\
&+& \left.\frac{1}{h v_1}\left(|t_{13}|^2+|t_{14}|^2+ \cdots \right)\right. \nn \\
&+& \left.\frac{1}{h v_2}\left(|t_{23}|^2+|t_{24}|^2+ \cdots \right) \right] \nn
\eea
Due to time reversal symmetry, ${\tilde r_{12}}={\tilde r_{21}}$ \& ${\tilde t_{12}}={\tilde t_{21}}$.\\
Inside the square bracket $\displaystyle[\hspace{.2cm}]$,\\
in the 3rd term we put ${\tilde r_{12}}={\tilde r_{21}}$, \\
in the 4th term we put ${\tilde r_{21}}= {\tilde r_{12}}$,\\
in the 7th term we put ${\tilde t_{12}}={\tilde t_{21}}$, \\
in the 8th term we put ${\tilde t_{21}}= {\tilde t_{12}}$.
\bea
\mbox{Therefore},\nn\\
\rho (E) &=& 2\left[\frac{1+|{\tilde r_{11}}|^2+|{\tilde r_{21}}|^2}{h v_1}\int_{-\infty}^0 dx \right.\nn\\
&+&\left. \frac{1+|{\tilde r_{12}}|^2+|{\tilde
r_{22}}|^2}{h v_2}\int_{-\infty}^0 dx \right.\nn \\
& + &\frac{|{\tilde t_{11}}|^2+|{\tilde t_{21}}|^2}{h v_1}\int_0^{\infty}dx
\hspace{.02cm}+\hspace{.02cm}\frac{|{\tilde t_{12}}|^2+|{\tilde t_{22}}|^2}{h v_2}\int_0^{\infty}dx\nn \\
& + &\frac{2|{\tilde r_{11}}|}{h v_1}\int_{-\infty}^0 dx \hspace{.2cm}Cos(2k_1x+\eta_1)\nn \\
&+&\frac{2|{\tilde r_{22}}|}{h v_2}\int_{-\infty}^0 dx \hspace{.2cm}Cos(2k_2x+\eta_2)\nn \\
& + & \frac{1}{h v_1}\left(|t_{13}|^2\hspace{.02cm}+\hspace{.02cm}|t_{14}|^2 \hspace{.02cm}+ \cdots \right)\nn \\
& + &\left. \frac{1}{h v_2}\left(|t_{23}|^2\hspace{.02cm}+\hspace{.02cm}|t_{24}|^2\hspace{.02cm}+ \cdots \right)\right]\nn
\eea
Now adding and subtracting the following terms inside $\displaystyle[\hspace{.2cm}]$,\\
\hfill\\
$\displaystyle{\frac{|{\tilde t_{11}}|^2}{h v_1}\int_{-\infty}^0 dx}$, 
$\displaystyle{\frac{|{\tilde t_{21}}|^2}{h v_1}\int_{-\infty}^0 dx}$, 
$\displaystyle{\frac{|{\tilde t_{22}}|^2}{h v_2}\int_{-\infty}^0 dx}$, 
$\displaystyle{\frac{|{\tilde t_{12}}|^2}{h v_2}\int_{-\infty}^0 dx}$,
\bea
\mbox{we get},\nn\\
\rho (E) &=& 2\left[\frac{1 + |{\tilde r_{11}}|^2 + |{\tilde r_{21}}|^2 + |{\tilde t_{11}}|^2 + |{\tilde t_{21}}|^2}{h v_1}\int_{-\infty}^0 dx \right. \nn\\
& + & \frac{1+|{\tilde r_{12}}|^2+|{\tilde r_{22}}|^2+|{\tilde t_{12}}|^2+
|{\tilde t_{21}}|^2}{h v_2}\int_{-\infty}^0 dx\nn\\
& + &\frac{2|{\tilde r_{11}}|}{h v_1}\int_{-\infty}^0 dx \hspace{.2cm}Cos(2k_1x+\eta_1)\nn\\
& + &\frac{2|{\tilde r_{22}}|}{h v_2}\int_{-\infty}^0 dx \hspace{.2cm}Cos(2k_2x+\eta_2)\nn \\
& + & \frac{1}{h v_1}\left(|t_{13}|^2\hspace{.02cm}+\hspace{.02cm}|t_{14}|^2 \hspace{.02cm}+ \cdots \right)\nn \\
& + & \left. \frac{1}{h v_2}\left(| t_{23}|^2\hspace{.02cm}+\hspace{.02cm}|t_{24}|^2\hspace{.02cm}+ \cdots \right) \right]\nn
\eea

Now \hspace{.2cm}$|{\tilde r_{11}}|^2 + |{\tilde r_{21}}|^2 + |{\tilde t_{11}}|^2 + |{\tilde t_{21}}|^2=1$ \\

and \hspace{.2cm}$|{\tilde r_{12}}|^2+|{\tilde r_{22}}|^2+|{\tilde t_{12}}|^2+
|{\tilde t_{21}}|^2=1$ \\
\bea
\mbox{Thus},\nn\\
\rho (E) & = & \frac{2}{h v_1}\int_{-\infty}^{\infty}dx \hspace{.2cm} + \hspace{.2cm}\frac{2}{h v_2}\int_{-\infty}^{\infty}dx\nn\\
& + &\frac{2|{\tilde r_{11}}|}{h v_1}\int_{-\infty}^{\infty} dx
\hspace{.2cm}Cos(2k_1x+\eta_1)\nn\\
& + &\frac{2|{\tilde r_{22}}|}{h v_2}\int_{-\infty}^{\infty} dx \hspace{.2cm}Cos(2k_2x+\eta_2)\nn \\
& + &\frac{2}{h v_1}\left(|t_{13}|^2\hspace{.02cm}+\hspace{.02cm}| t_{14}|^2 \hspace{.02cm}+ \cdots \right)\nn \\
& + &\frac{2}{h v_2}\left(t_{23}|^2\hspace{.02cm}+\hspace{.02cm}|t_{24}|^2\hspace{.02cm}+ \cdots \right)\hspace{2.0cm}\mbox{(iii)}\nn
\eea
Now $\displaystyle \frac{2}{h v_1}\int_{-\infty}^{\infty}dx + \frac{2}{h v_2}\int_{-\infty}^{\infty}dx = \rho_0 (E)$ i.e. DOS in the absence of scatterer. 
Also according to some text books \cite{gha},
for plane wave states the current conservation condition
$$\frac{\pa}{\pa t}\int_{\Omega} \psi^*\psi \hspace{.2cm}d\tau \hspace{.2cm}+\hspace{.2cm}\int_{S} \vec J.\hat n \hspace{.2cm}ds \hspace{.2cm}=\hspace{.2cm}0$$
can be satisfied if and only if the wave function vanishes at $\pm \infty$ for
all energy. Thus the 3rd and 4th term of Eq.(iii) of this 
section is $0$. Besides, neglecting these oscillatory
terms in the leads is very standard in the derivation of Friedel
sum rule \cite{yeycm}, where it is assumed that the carrier
concentration is so high in the leads that the leads
are non-polarizable. However, assuming non-polarizable leads
alone does not eradicate all energy dependence of the self energy
as thought in Ref. \cite{yeycm} (see Eqs. 2 and 6 therein, the
parameter $t_\alpha$ can be strongly energy dependent, which
is dispersive behavior). 
So we get\\
\bea
\rho (E) & = &\rho_0 (E)\hspace{.2cm}+\hspace{.2cm}\frac{2}{h v_1}\left(|t_{13}|^2\hspace{.2cm}+\hspace{.2cm}|t_{14}|^2 \hspace{.2cm}+ \cdots \right)\nn\\
&+&\frac{2}{h v_2}\left(|t_{23}|^2\hspace{.2cm}+\hspace{.2cm t_{24}}|^2\hspace{.2cm}+ \cdots \right)\nn
\eea
\bea
\mbox{Thus} \hspace{.4cm} \rho(E)-\rho_0(E)&=& \frac{2}{h v_1}\left(|t_{13}|^2\hspace{.2cm}+\hspace{.2cm}|t_{14}|^2 \hspace{.2cm}+ \cdots \right)\nn\\
&+&\frac{2}{h v_2}\left(|t_{23}|^2\hspace{.2cm}+\hspace{.2cm t_{24}}|^2\hspace{.2cm}+ \cdots \right)\nn\\
&=& \left(\rho-\rho_0\right)_R\nn
\eea

\end{multicols}


\begin{thebibliography}{99}
\bibitem{com} From the derivation of the FSR (one can see the book by Ziman
in Ref. \cite{fri52}), that the scattering phase shift is defined such
that when we set the scattering potential to 0 the scattering
phase shift is also 0. However, in the double delta function potential
the $arg(t)=kl$ when the strength of the delta function potentials
are set to zero. Hence the scattering phase shift is $arg(t)-kl$.
However, the term coming from the derivative of $kl$ cancels with the
term coming from $\rho_0$, leading to the Eq. in (11).
\end{thebibliography}
\end{document}